\documentclass[a4paper]{article}
\usepackage{graphicx}
\usepackage{amsmath,amssymb,amsfonts}
\title{Transition to turbulence in wall-bounded flows:\\
Where do we stand?\\
{\large Paul MANNEVILLE}\\[1ex]
{\normalsize Hydrodynamics Laboratory, \'Ecole Polytechnique, Palaiseau, 91128, France\\[-2ex]
\tt paul.manneville@ladhyx.polytechnique.fr}
}
\date{to appear in {\sl Mechanical Engineering Review}, Vol. 3, No. 2, July 1st, 2016.}

\def\BM{\begin{displaymath}}
\def\EM{\end{displaymath}}
\def\BE{\begin{equation}}
\def\EE{\end{equation}}
\def\BA{\begin{eqnarray}}
\def\EA{\end{eqnarray}}
\def\BAN{\begin{eqnarray*}}
\def\EAN{\end{eqnarray*}}

\def\RE{\ensuremath{R}}
\def\bnabla{\mbox{\boldmath{$\nabla$}}}

\begin{document}

\maketitle

\begin{abstract}
In this essay, we recall the specificities of the transition to turbulence in wall-bounded flows and present recent achievements in the understanding of this problem.
The transition is abrupt with laminar-turbulent coexistence over a finite range of Reynolds numbers, the transitional range.
The archetypical cases of Poiseuille pipe flow and plane Couette flow are first reviewed at the phenomenological level, together with a few other flow configurations.
Theoretical approaches are then examined with particular emphasis on the existence of special nontrivial solutions to the Navier--Stokes equations at finite distance from laminar flow.
Dynamical systems theory is most appropriate to analyze their role, in particular with respect to the transient character of turbulence in the lower transitional range.
The extensions needed to deal with the prominent spatiotemporal features of the transition are then discussed.
Turbulence growth$\mskip1mu/\mskip1mu$decay in terms of statistical physics of many-body systems and the relevance of directed percolation as a stochastic process able to account for it are next scrutinized.
To conclude, we advocate the recourse to well-designed modeling able to provide us with a conceptually coherent picture of the full transitional range and put forward some open issues.\\[1ex]

{\bf keywords}: Transitional Flow, Pipe Flow, Channel Flow, Dynamical Systems, Chaos, Turbulence.
\end{abstract}

\section{Context and issues\label{S1}}
Turbulent shear flows have transport properties that are considerably enhanced when compared to their laminar counterparts.
It is therefore of primary interest to us to understand how turbulence develops upon increasing the shearing rate.
The first approaches to this problem  date back to the last quarter of the XIXth century with the work of Reynolds on the transition in pipe flow, coming after the first formal studies of flow stability by Helmholtz, Kelvin, Rayleigh,  ca.~1870--1880 and later by Orr and Sommerfeld, ca. 1908 (Eckert, 2010; Eckhardt ed., 2009).
Mostly dealing with mathematically infinitesimal perturbations and focusing on linear normal-mode dynamics, the conventional stability theory (Lin, 1955) experiences difficulties in accounting for the emergence of turbulent behavior directly out of laminar flow, as anticipated by Reynolds (1883) who understood the importance of finite-amplitude localized disturbances.
It was not until the rise of nonlinear dynamics and chaos theory in the 1970's (Aubin \& Dalmenico, 2002),  the non-modal approach to transient growth (Trefethen {\it et al.}, 1993; Grossmann, 2000), and the systematic  development of numerical simulations of the Navier--Stokes equations in relevant flow configurations (Moin \& Mahesh, 1998), that some theoretical understanding has been gained out of the accumulation of phenomenological evidence on the direct transition {\it via\/} turbulent {\it puffs\/} or {\it slugs\/} in pipe flow (Reynolds, 1883; Lindgren, 1951; Wygnanski \& Champagne, 1973; {\it etc.}) and, in other flows, {\it  spots\/} (Emmons, 1951; Carlson {\it et al.}, 1982; {\it etc.}),  {\it spirals\/} or {\it oblique bands\/} (Coles, 1962; Andereck {\it et al.}, 1986; Prigent, 2001; {\it etc.}).
Illustrative examples of these transitional structures will be given later on.

Before briefly presenting experimental results in more detail, let us continue to set the frame and point out the origin and nature of the difficulty of the problem at stake: Incompressible flows of simple fluids are governed by the Navier--Stokes (NS) equation,
$\partial_t\mathbf v+\mathbf v\cdot\bnabla\mathbf v=-\bnabla p+\RE^{-1} \bnabla^2\mathbf v$,
and the continuity condition $\bnabla\cdot \mathbf v=0$. The Reynolds number \RE\ is the control parameter.
It compares the shearing rate $\delta{\mskip-1mu} U/\delta{\mskip-1mu} L$ and the viscous dissipation rate $\nu/\delta{\mskip-1mu} L^2$, namely $\RE=(\delta{\mskip-1mu} U/\delta{\mskip-1mu} L)/ (\nu/\delta{\mskip-1mu} L^2)=\delta{\mskip-1mu} U\delta{\mskip-1mu} L/\nu$.
In these expressions $\delta{\mskip-1mu} U$ measures the typical velocity variation over the characteristic length scale $\delta{\mskip-1mu} L$ of the considered fluid configuration, and quantity $\nu$ is the kinematic viscosity.
When $\RE\ll 1$, the viscous dissipation rate is large and viscosity keeps the flow {\it laminar} (smooth), while in the opposite situation, $\RE\gg1$, spatiotemporally irregular flow, currently termed  {\it turbulent},  prevails in practice.
A sketch of the general setting of the transition to turbulence is given in~Fig.~1.

In view of its importance in situations of engineering interest, the transitional regime from laminar flow to turbulence has been mainly studied as a {\it local\/} stability problem, i.e. against infinitesimal perturbations, by solving the eigenvalue problem for the NS equation linearized around the base flow solution.
That analysis possibly yields a threshold Reynolds number $\RE_{\rm c}$ beyond which the flow is {\it unconditionally unstable\/}, i.e. always departs from the base profile (Joseph, 1976), see Fig.~1(a).
As is well known, two main cases can be identified depending on whether the base flow displays an inflection point or not, Fig.~\ref{F2}, as reviewed by Huerre and Rossi (1998).
A flow with velocity profile displaying an inflection point with a vorticity maximum may experience a mechanical instability of Kelvin--Helmholtz type that produces spanwise vorticity at rather low $\RE$ whereas viscosity plays its expected mitigating role.
As $\RE$ is increased further, these vortical structures rather rapidly degenerate into turbulence according to a `globally supercritical' scenario, with the meaning of a progressive development of  a spatiotemporally complicated and unpredictable flow pattern.
\begin{figure}[t]
\begin{center}
\includegraphics[width=0.9\textwidth]{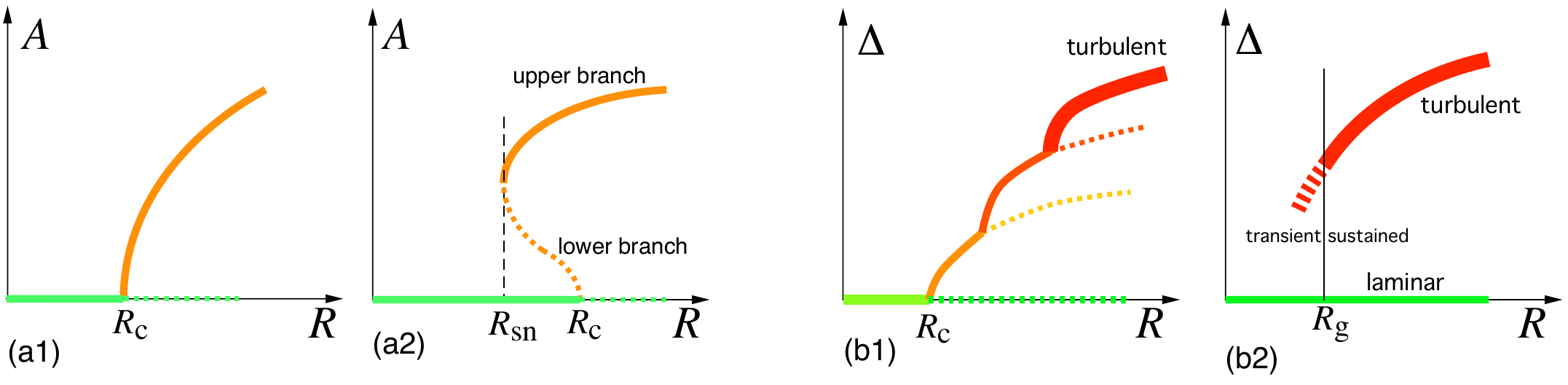}
\end{center}

\caption{ \label{F1} Local {\it versus\/} global. (a) Standard bifurcation theory deals with dynamical behavior changes that can be characterized by a few amplitudes, viz. $A$, as functions of control parameters, viz. $R$. `Local' means local in phase space, where the new state $A\ne0$ remains `close' to the base state $A=0$. The bifurcation can be supercritical (a1) or subcritical (a2). The linear instability threshold is $R_{\rm c}$ and the base state is unconditionally unstable for $R\ge R_{\rm c}$. When subcritical, the bifurcation typically takes place through a `saddle-node' at some threshold $R_{\rm sn}$ beyond which new states, one stable (`upper-branch', continuous line) and one unstable (`lower-branch', dotted line) coexist with the still linearly stable base state. (b)~Transition to turbulence generally takes place through scenarios that develop over wider ranges of control parameter and involve larger regions of phase space. The distance to base state is now denoted $\Delta$ without reference to a specific bifurcation amplitude. The `globally supercritical' scenario refers to a cascade of bifurcations where the bifurcating state replaces the bifurcating state and remains close to it, hence an overall continuous scenario (b1). In contrast, the `globally subcritical' case is fully discontinuous~(b2). The base state is seen to coexist with a branch of new solutions far from it, here represented by a thick line, with the meaning that $\Delta$ is only the statistical mean of a highly fluctuating quantity. Generically, this new branch can be followed at decreasing $\RE$ and a transition is observed between a sustained regime and a regime where the base state is recovered at the end of  long transient. The global stability threshold $\RE_{\rm g}$ is precisely defined as the value of $\RE$ below which the base state is the only asymptotic state in the long term whatever the size of the perturbation brought to it. The main difficulty of this case is to understand the nature of the nontrivial states far from the base state, and to know how to reach them~(\S\ref{S3}). Another question is whether and in which sense the transition is discontinuous, i.e. whether $\Delta$ is finite at $\RE_{\rm g}$ or sustained turbulent states can be maintained with arbitrarily small $\Delta$ just above $\RE_{\rm g}$ (\S\ref{S4}).}.
\end{figure}
Examples of globally supercritical scenarios are given by the mixing layer, the jet, or the wake of a bluff body.
In contrast, non-inflectional base flow profiles are typically stable against mechanical instabilities at transitional values of $\RE$.
Viscous effects  play a more important role and, in particular, push the essential steps of the transition to turbulence at larger $\RE$.
First, viscosity controls the establishment of the base profile, parabolic Hagen--Poiseuille  in a cylindrical pipe submitted to constant mass flux or constant pressure gradient  (HPF in the following), linear in the flow between plates differentially driven at different speed (plane Couette flow, PCF), parabolic in a rectangular channel (plane Poiseuille flow, PPF),  Blasius in a boundary layer along a flat wall  (BBL), {\it etc}.
Next, being ultimately responsible for the development of Tollmien--Schliching (TS) waves, viscous effects play a much nastier role; see (Gersten, 2009) for a historical perspective, and (Schmid \& Henningson, 2001) for a recent technical account.
Linear and nonlinear contributions to the dynamics in fact compete at intermediate values of $\RE$, not high enough to make the concept of  {\it  developed turbulence\/} relevant, but high enough for permitting the existence of nontrivial flow regimes away from laminar flow, though well below the threshold $\RE_{\rm c}$ for TS waves.
For example,  $\RE_{\rm c}=5772$ in PPF (Orszag, 1971) whereas turbulent spots were found to develop around $\RE\sim 1000$ (Carlson {\it et al.}, 1982), and a mildly turbulent state in the form of oblique turbulent bands can even be maintained down to $\RE_{\rm g}$ less than $800$ (Tuckerman {\it et al.}, 2014; Tsukahara \& Ishida, 2014).
Here subscript `g' stands for `global', the {\it global stability threshold\/} being the value of $\RE$ below which laminar base flow remains {\it unconditionally stable}, i.e. whatever the shape and amplitude of disturbances (Joseph, 1976), see Fig.~1(b2).
Linear instabilities may even not exist, i.e. $\RE_{\rm c} = \infty$.
For instance, PCF is linearly stable for all $\RE$ (Romanov, 1973) whereas $\RE_{\rm g}\simeq 325$ (Bottin {\it et al.}, 1998) and similarly for HPF (Salwen {\it et al.}, 1980; Meseguer \& Tretheten, 2003) with $\RE_{\rm g}\simeq 2040$ (Avila {\it et al.}, 2011).
\begin{figure}[t]
\begin{center}
\includegraphics[height=0.22\textwidth]{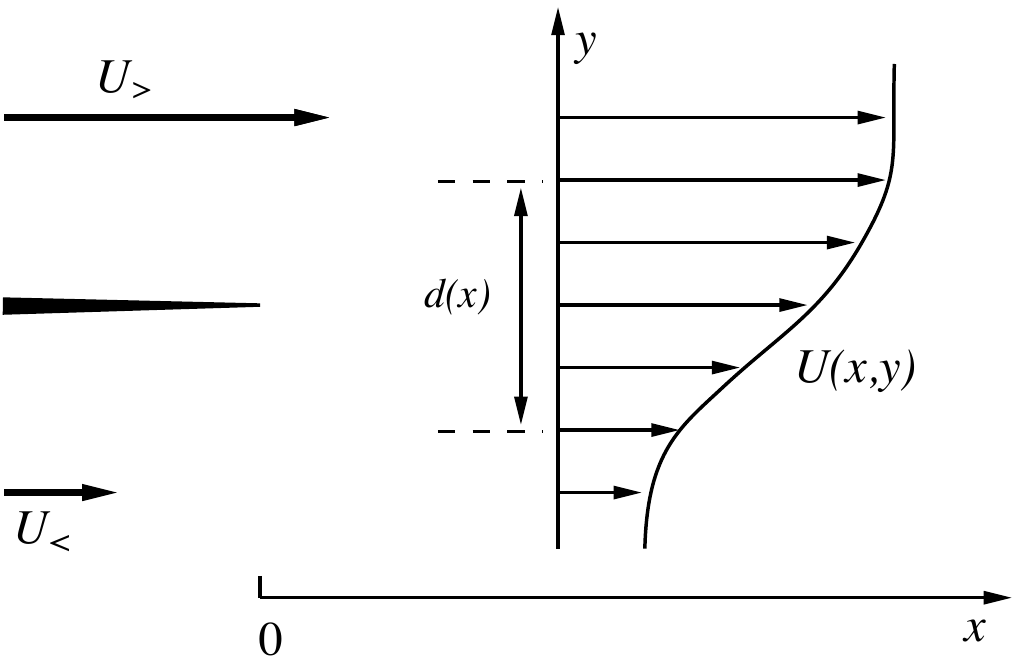}\hskip6em
\includegraphics[height=0.22\textwidth]{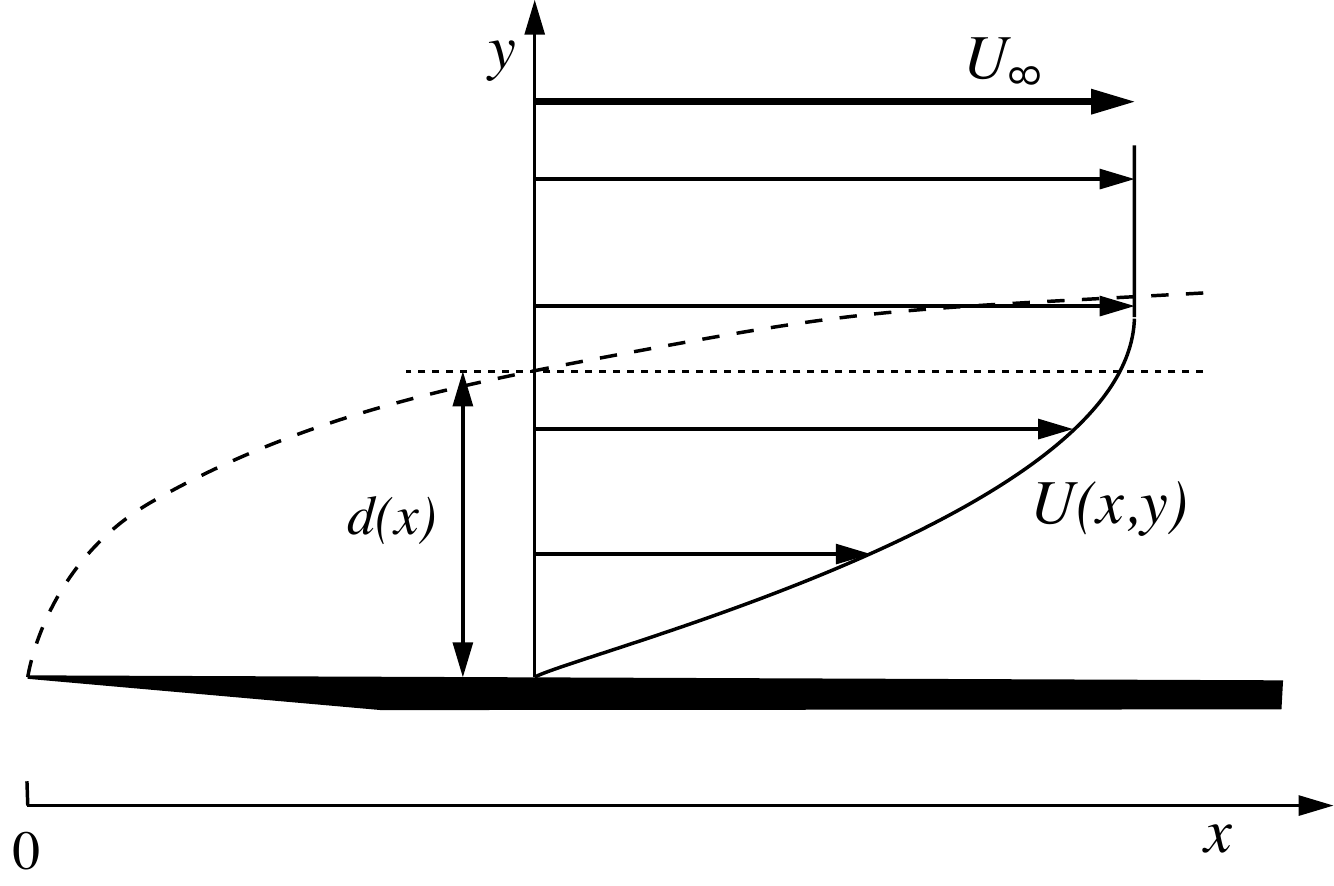}
\end{center}
\caption{\label{F2} The mixing layer developing downstreams a splitter plate is prototypical of flow with an inflectional base flow velocity profile (left). The Blasius boundary layer flow along a plate is an example of non-inflectional wall-bounded flows (right). These laminar profiles evolve in space  as functions of the distance to the edges under viscous diffusion, in contrast with Hagen--Poiseuille (pipe) flow driven by a constant pressure gradient (Fig.~\ref{F3}) or plane simple-shear Couette flow (Fig.~\ref{F4}).}
\end{figure}
In the following we first present  the phenomenology of these two most extreme cases in reverse order, HPF (\S\ref{S2.1}), next PCF (\S\ref{S2.2}).
They mainly differ in the effective physical dimension of their respective geometries, one-dimensional (1D) with downstream transport along the tube axis for HPF, or 2D in the plane of the driving plates and no advection for PCF.
A few other flow configurations will also be more sketchily examined (\S\ref{S2.3}).
Our current theoretical understanding of the processes underlying the transition will be 
considered afterwards in Sections \ref{S3} and \ref{S4}.

The bibliography on the transition in wall-bounded flows is plethoric.
Here the (already abundant) literature cited is kept at a minimum, with only some of what we believe to be the most significant original contributions.
Further references can be found in the books, compilations, and review articles listed.
Sections~\ref{S2} and \ref{S3} below may be viewed as a summary of the review part of our article (Manneville, 2015a).
Section~\ref{S4} is focused on issues of recent interest. 


\section{Phenomenology\label{S2}}

In the range of Reynolds numbers where the transition to turbulence is observed, the flows of interest are linearly stable.
The transition {\it to\/} turbulence shows great sensitivity to the exact {\it shape\/} and {\it amplitude\/} of perturbations that induce it for $R$ large enough. 
The extension of the attraction basin of laminar flow is then under scrutiny and one traditionally starts from a base flow with some level of residual turbulence and$\mskip0.5mu/\mskip0.5mu$or additionally introduces spatially localized disturbances of finite amplitude.
Marked hysteresis is also observed since turbulent flow -- sustained or not this will be an important question -- is currently observed in the transitional range.
The transition {\it from\/} turbulence as  $\RE$ is decreased is therefore also of interest, again sensitive to the protocol used for changing $\RE$, mainly sudden {\it quenches\/} from high to low values of $\RE$ or, on the contrary, slow regular decrease termed {\it annealing}.

\subsection{Hagen--Poiseuille flow\label{S2.1}}

\begin{figure}[t]
\begin{center}
\includegraphics[height=0.1\textwidth]{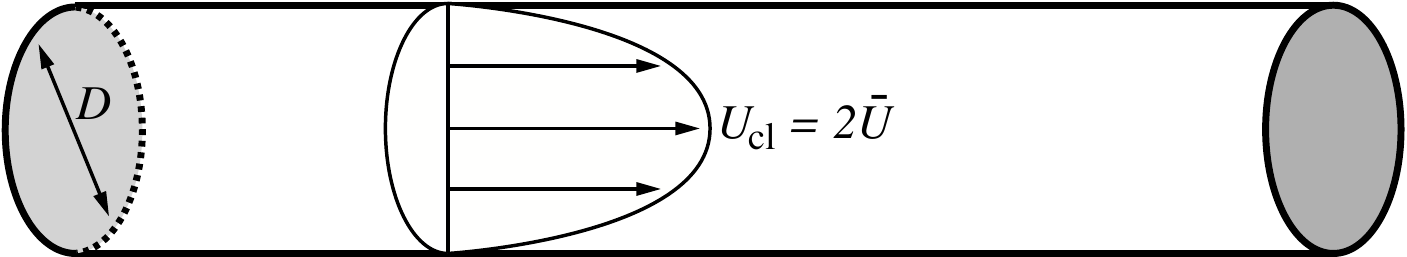}\\[2ex]
\includegraphics[width=0.8\textwidth,height=0.1\textwidth]{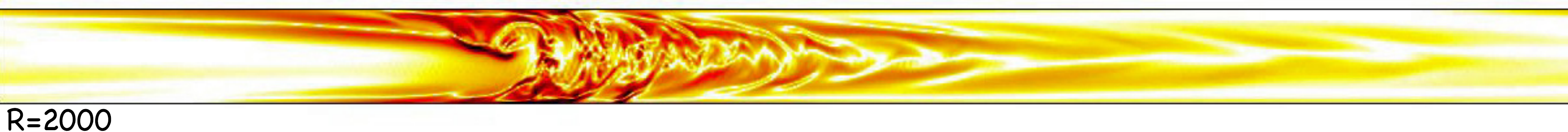}\\
\includegraphics[width=0.8\textwidth,height=0.1\textwidth]{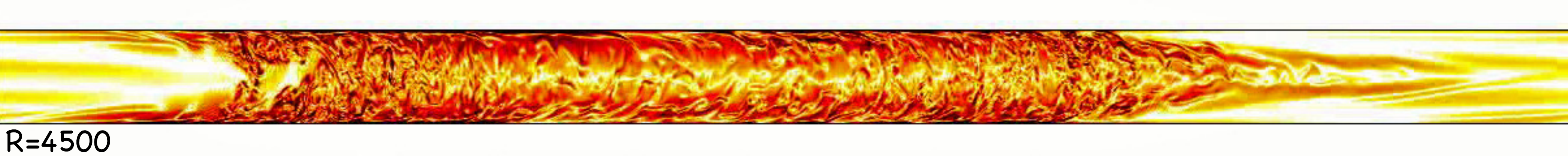}
\end{center}
\caption{\label{F3} Top: Sketch of the pipe-flow experiment: A constant pressure difference can be applied between the inlet and the outlet of the pipe, or else a constant mass flux can be produced by withdrawing a piston at constant velocity $\bar U$, sucking the fluid from a reservoir.
Bottom: Typical aspect of puffs and slugs in transitional HPF: Equilibrium puff at $R=2000$ and growing slug at $R=4500$.
Snapshots from numerical simulations by Duguet {\it et al.}  (2010b), courtesy: Y. Duguet.
The flow is from left to right.
Notice the similarly sharp laminar-turbulent trailing edge (upstream, on the left) in both cases.
In contrast, the turbulent-laminar leading edge (downstream, right) is more abrupt for the slug than for the puff.
 Owing to its larger $\RE$, the slug also has a finer, vigorously fluctuating,  internal structure than the puff that displays a more coherent, only mildly chaotic, internal structure.}.
\end{figure}

The flow through a tube (HPF), encountered in ordinary hydraulic context, from water distribution networks to pipelines, is apparently easily implemented, but achieving a near-ideal setup appropriately dealing with length effects is not an easy task since one has to cope with downstream advection along the tube (streamwise 1D system).
The different flow regimes developing in a tube of circular section were first studied by Reynolds (1883) who already observed that the parabolic base flow can be maintained up to high values of a control parameter he identified, the {\it Reynolds number} introduced above.
For intermediate values of $\RE$ he noticed brief episodes of irregular {\it sinuous\/} motion, that he called {\it flashes of turbulence\/}, while the flow was becoming more uniformly turbulent when $\RE$ increased.
Here $\RE$ is constructed with the mean speed $\bar U$ as $\delta{\mskip-1mu} U$ and the tube's diameter $D$ as $\delta{\mskip-1mu} L$ (or equivalently the laminar centerline velocity $U_{\rm cl}=2\bar U$ and the tube's radius $D/2$, see Fig.~\ref{F3}, top).
His findings are fully consistent with modern results even quantitatively since, with a setup that can still be run in Manchester, he located the transition around $\RE\simeq2000$ and was able to maintain laminar flow much beyond $10^4$. A strict mathematical proof that the base flow is linearly stable for all $\RE$ seems still lacking but the analysis of Salwen {\it et al.} (1981) and the computations of Meseguer \& Tretheten (2003) strongly suggest that it is.
[A similar situation holds for the flow in a pipe of square to slightly rectangular section. As shown by Tatsumi \& Yoshimura (1990),  when the channel's width-to-height ratio is larger than 3.2, $\RE_{\rm c}$ decreases from infinity to a finite value that tends to Orszag's $\RE_{\rm c}=5772$ as the width further increases.]

Experimental studies of HPF have been numerous, only a few markers are given here; see (Mullin, 2011) for a review of experimental results and (Willis {\it et al.}, 2008) or (Eckhardt ed., 2009) for more emphasis on theoretical and numerical issues.
An important work has been performed by Lindgren starting with his PhD thesis (1957) and extending over the decade that followed.
Studying Reynolds' flashes he carefully measured the speeds of their leading and trailing edges and identified a regime of invasive turbulent plugs that he called {\it slugs\/}.  
These early results have been nicely reviewed by Coles (1962) who also considered other important wall-bounded configurations, in particular presenting his own findings on circular Couette flow (see later).
In a second instance comes the work of Wygnanski and coworkers (1973, 1975) who isolated a separate regime in the lowest part of the transitional range where the flashes take the form of localized mildly turbulent localized structures they called {\it puffs\/}.
With the stimulus of advances in nonlinear dynamics, the most significant recent achievements relate to the lowest part of the transitional range and a reliable localization of the global stability threshold $\RE_{\rm g}$, as a followup of studies by Mullin and co-workers (Darbyshire \& Mullin, 1995; Hof {\it et al\/}, 2003; review by Mullin, 2011).

Turbulent puffs are coherent structures of roughly constant length with fuzzy arrow-shaped heads and steep trailing edges, in contrast with expanding slugs with more abrupt and fast moving laminar-turbulent interfaces (Wygnanski {\it et al.}, 1973; Nishi {\it et al.}, 2008; Duguet {\it et al.}, 2010b; Barkley {\it et al}, 2015).
 See Fig.~\ref{F3} for illustrations at $\RE=2000$ and $\RE=4500$, respectively.
They were long thought to be steady-state or {\it equilibrium\/} structures (Wygnanski {\it et al.}, 1975) but found to be long-lived {\it transients\/} decaying to laminar flow on the low-$\RE$ side of the transitional range (Darbyshire \& Mullin, 1995).
These transients were found with an exponential distribution of lifetimes $\tau_{\rm d}$, `d' for `decay', and the mean lifetime $\bar\tau_{\rm d}$ seen to increase with $\RE$.
Whether or not $\bar\tau_{\rm d}(\RE)$ actually diverged as $\RE$ approached some critical value  long stayed a matter of debate but the consensus is now of a double-exponential growth $\bar\tau_{\rm d}(\RE) = \exp[\exp(a_{\rm d}\RE - b_{\rm d})]$ (Hof {\it et al.\/}, 2006, 2008), which we prefer write $\tau_{\rm d}(\RE) = \exp\{\exp[b_{\rm d}(\RE/\RE_{\rm d} -1)]\}$ with $b_{\rm d}\approx8.5$ and $\RE_{\rm d}\approx1530$, so that $\RE_{\rm d}$ presents itself as the value of $\RE$ at which decay would be ``immediate,'' i.e. of the order of the turnover time $D/\bar U$, whereas $b_{\rm d}$ is a number of order unity.

When $\RE$ is becoming larger, turbulence may be expected to become sustained, which is actually the case through a new process:
Besides {\it puff decay\/} that kills localized turbulence, {\it puff splitting\/}, in which a mother puff generates a newborn ahead of it, helps propagating turbulence (Nishi {\it et al.}, 2008; Avila {\it et al.}, 2011).
Puff-splitting events were found to occur at random time intervals that are also exponentially distributed with another characteristic time $\tau_{\rm s}$, `s' for `splitting'.
Turbulence is still weak in the sense that it is intense inside the puffs but intermittent in space at given $\RE$.
As $\RE$ increases, the turbulence level increases as splitting events become more frequent, with a mean waiting time also varying super-exponentially with $\RE$: $\bar\tau_{\rm s}(\RE)=\exp[\exp(-a_{\rm s}\RE + b_{\rm s})]$ (Avila {\it et al}, 2011), which we write $\bar\tau_{\rm s}(\RE)=\exp\{\exp[ b_{\rm s} (1-\RE/\RE_{\rm s})]\}$ with $b_{\rm s}\approx9.16$ and $\RE_{\rm s}\approx2940$, the $\RE$ value at which splitting would be ``immediate.''
The global stability threshold can then be fixed by stating that splitting compensates spontaneous decay at the statistical level: $\bar\tau_{\rm s}(\RE_{\rm g}) = \bar\tau_{\rm d}(\RE_{\rm g})$,
yielding $\RE_{\rm g}\approx 2040$ (Avila {\it et al}, 2011).
The detailed physics behind decay or splitting is still not well understood (Shimizu {\it et al.}, 2014).
Fluid particles completely pass through the puffs that are transported downstream slightly slower than the mean flow.
As $\RE$ further increases, the turbulent/laminar interfaces change their structures and puffs turn into more aggressive slugs with leading edges propagating faster downstream and trailing edges able to propagate upstream (Duguet {\it et al.}, 2010b).
The most recent study by Barkley {\it et al.} (2015) indicates a behavior change of these interfaces at $\RE \approx 2250$ with intermittent laminar pockets up to $\RE\approx 3000$ and essentially uniform turbulence above.  
Laminar intervals being progressively nibbled, a more-or-less uniformly turbulent state is then expected sufficiently far downstream.

\subsection{Plane Couette flow\label{S2.2}}

Plane Couette flow (PCF), the simple shear flow developing between two infinite counter-sliding parallel plates is our second example.
Being linearly stable flow for all $\RE$ (Romanov, 1973), it is the simplest representative of planar flows experiencing a direct transition to turbulence due to finite amplitude perturbations.
Owing to the existence of a relevant spanwise direction in addition to the streamwise direction, these systems are effectively two-dimensional (2D) and the laminar-turbulent coexistence organizes itself in more or less regular domains separated by fluctuating interfaces.
In PCF, the lack of downstream transport (PPF, BBL) and overall streamwise space dependence (BBL) makes it in principle easy to study the transition in the long term but experimental difficulties in preparing good quality setups of sufficiently large aspect ratios may come and temper this expectancy.
\begin{figure}[t]
\begin{center}
\begin{minipage}{0.38\textwidth}
\includegraphics[width=\textwidth]{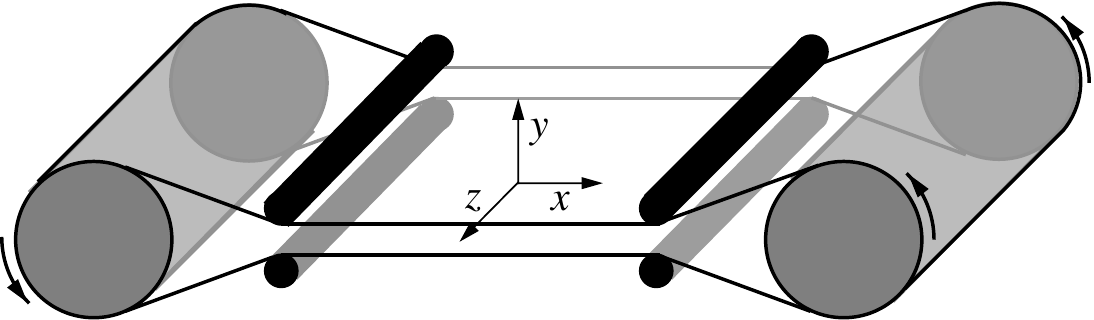}
\end{minipage}
 \hfill
\begin{minipage}{0.6\textwidth}
\includegraphics[width=\textwidth]{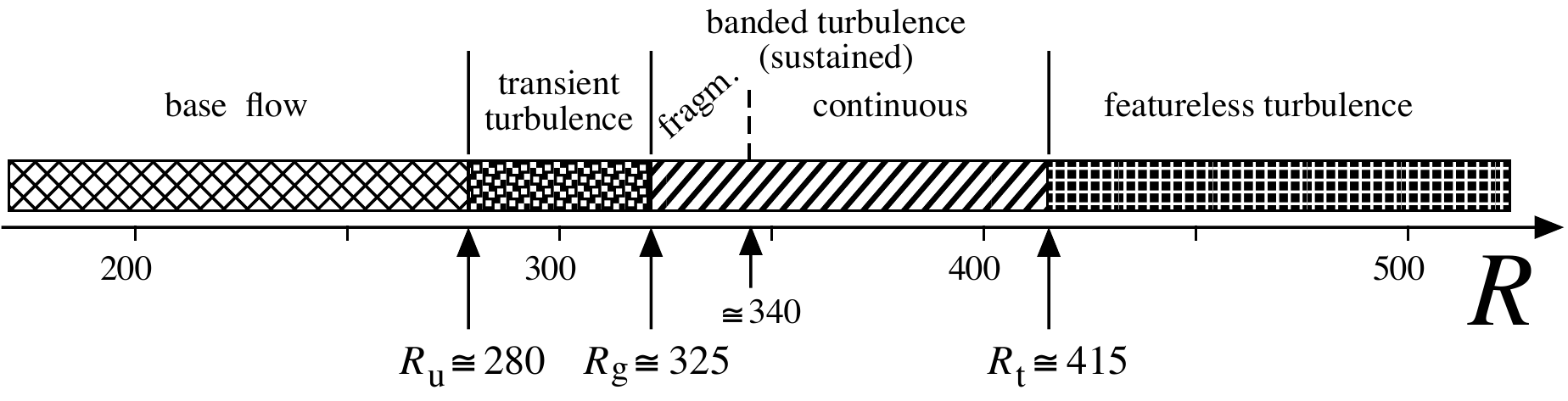}
\end{minipage}
\vspace{2ex}

\begin{minipage}{0.45\textwidth}
\includegraphics[width=\textwidth]{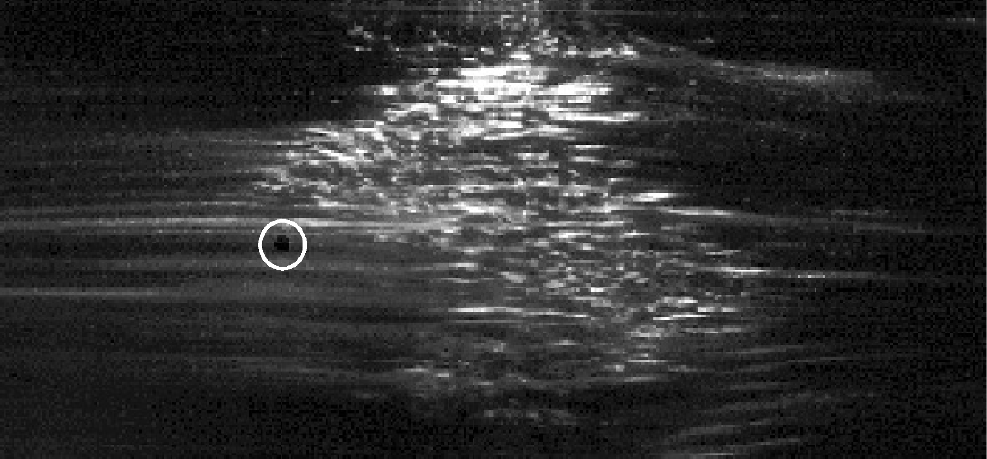}
\end{minipage}\hfill
\begin{minipage}{0.5\textwidth}
\includegraphics[width=\textwidth]{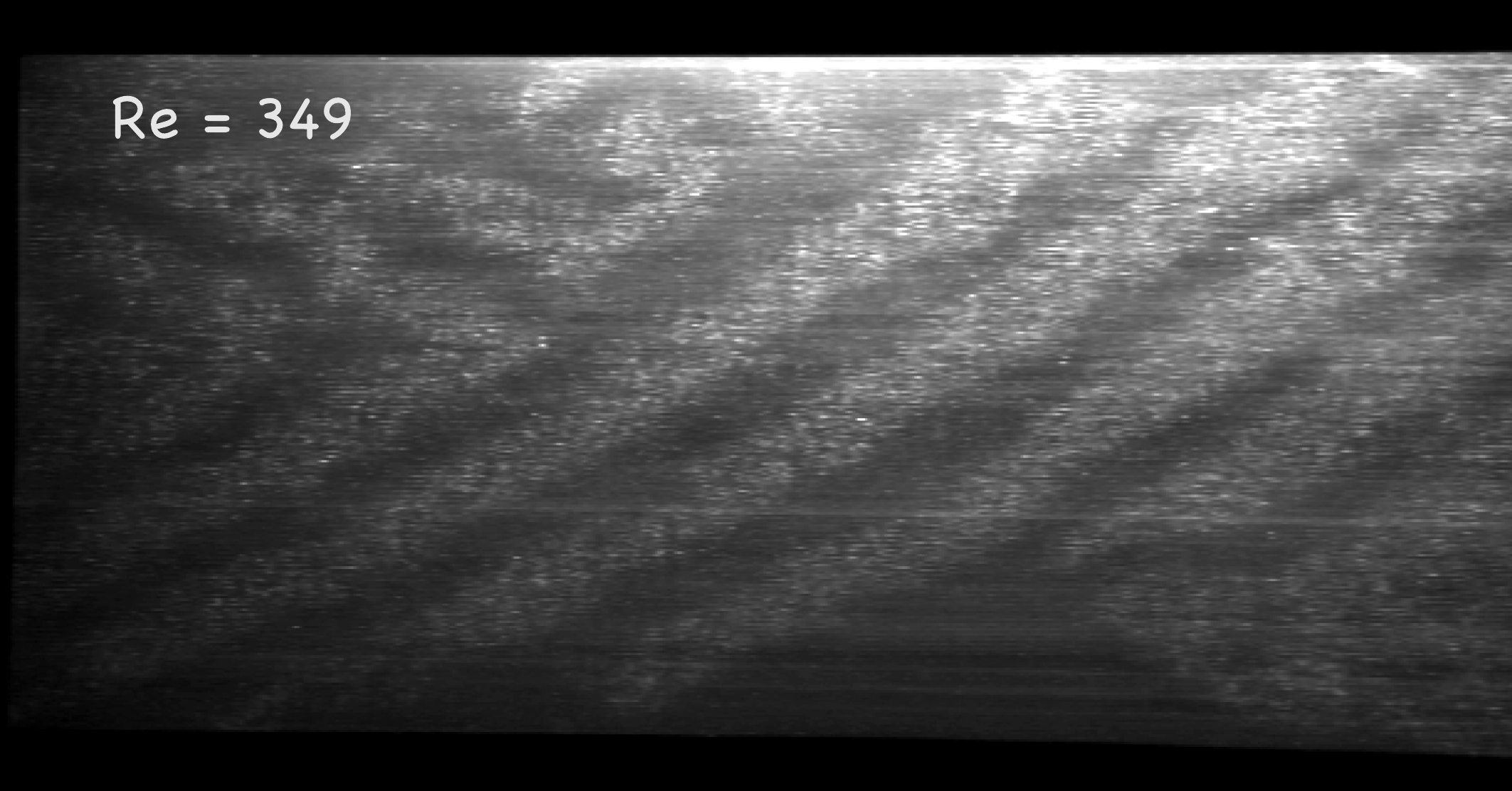}
\end{minipage}

\end{center}
\caption{\label{F4} Top: Sketch of the plane Couette flow set-up used at Stockholm and Saclay and bifurcation diagram as compiled from the work of the Saclay group (Bottin, 1998; Prigent, 2001).
Bottom, left: Mature turbulent spot slightly beyond $\RE_{\rm g}\approx 325$;
the white circle indicates the position of the initial localized perturbation;
the spot is slightly oblique with an irregular shape.
Courtesy: S. Bottin.
Bottom right: Turbulent bands at $\RE=349$.
The streamwise direction is horizontal.
Laminar flow is uniformly dark while turbulent flow scatters incident light and appears brighter. 
The two pictures have been reframed and are not at the same scale (original aspect ratios $140\times35$ vs. $335\times170$). The streamwise striations have spanwise widths of the same order of magnitude in both cases, though somewhat narrower in the bands owing to the larger value of $\RE$. Courtesy: A. Prigent.}
\end{figure}

Early experiments, e.g. by  Reichardt (1957), were mostly devoted, not to the transitional range, but  to the fully developed steady-state regime.
Mechanically simpler to produce,  the flow produced by a belt moving parallel to a wall at rest has a mean value that carries perturbations out the observation field like for HPF, PPF, BBL, \dots.
This drawback is absent from more recent setups developed in Stockhom, Sweden (Tillmark \& Alfredsson, 1992) and Saclay, France (Daviaud {\it et al.}, 1992) with a belt forming a closed loop driven by big cylinders and managing a gap of controlled width by four guiding rollers, see Fig.~4 (top, left).
The zero-mean flow condition is ensured by the two facing parts of the belt running at exactly the same speed but in opposite directions, which allowed to perform a detailed study of the emergence and sustenance of turbulence in the long term.

The Reynolds number is usually defined as $\RE=U_{\rm b}h/\nu$ where $U_{\rm b}$ is the linear speed of the band and $2h$ is the gap between the two faces of the band running parallel to each other. 
In addition the problem is controlled by the value of two aspect-ratios, $\Gamma_{x,z}:=L_{x,z}/2h$, where $L_x$ and $L_z$ are the streamwise and spanwise dimensions of the setup.
In the early 1990's, experiments were performed at moderate aspect-ratios, $\Gamma_x \sim 75$--$150$, $\Gamma_z \sim 35$.
Several devices were used to trigger turbulence (see the review by Prigent \& Dauchot, 2005), modifying the flow locally, either permanently or impulsively.
Tiny jets piercing the base flow profile during a short time interval were used as initial conditions (Tillmark \& Alfredsson, 1992; Daviaud {\it et al.}, 1992;  Dauchot \& Daviaud, 1995; Bottin {\it et al.}, 1998).
Such perturbations generate {\it turbulent spots\/} that grow or decay depending on the value of $\RE$.
Like the puffs, spots are transient below $\RE_{\rm g}$, with exponentially decreasing lifetime distributions and mean lifetime increasing with $\RE$, illustration in Fig.~4 (bottom, left).
The global stability threshold (against these specific perturbations) $\RE_{\rm g}$ was estimated first around $350$ and, in later experiments, somewhat lower, around $325$ as determined from lifetime measurements that suggested mean lifetimes diverging as $(\RE_{\rm g}-\RE)^{-1}$  (Bottin {\it et al.}, 1998), a behavior later questioned by the proponents of non-singular unlimited mean lifetime growth (Hof {\it et al.}, 2006).
Anyhow, aspect ratios were small, and lateral size effects presumably large, thus possibly biasing the lifetime estimates.

Mostly confirming the results while not making them more precise, experiments performed by Prigent (2001) at much larger aspect ratio ($335\times170$) brought interesting new features with respect to the spatial organization of the laminar and turbulent domains.
These experiment were conducted by starting from a uniformly turbulent flow at high $\RE$ and decreasing $\RE$ progressively.
They pointed out a second threshold $\RE_{\rm t}>\RE_{\rm g}$ below which an oblique modulation of the turbulence intensity developed (Fig.~4, bottom-right), with the modulation amplitude -- the corresponding Fourier mode intensity -- as the relevant {\it order parameter\/}.
This modulation takes the form of oblique bands alternatively laminar and turbulent with roughly constant streamwise wavelength $\lambda_x\approx 110h$ and spanwise wavelength $\lambda_z$ increasing from  $\approx 50h$ at large $\RE$ when the modulation appears around $\RE_{\rm t}\simeq 415$, to $\approx 85h$ when the pattern breaks down as $\RE$ approaches $\RE_{\rm g}\simeq325$.
The modulation amplitude was observed to decrease continuously to zero as a power law of the distance to $\RE_{\rm t}$ (Prigent {\it et al\/}., 2003).
This behavior is reminiscent of a standard forward bifurcation though ordinarily the base state is laminar rather than uniformly turbulent, and the control parameter decreased rather than increased.
The band pattern stayed steady and regular in the upper part of the range $[\RE_{\rm g},\RE_{\rm t}]$, but upon further decreasing $\RE$, around $\RE\approx 340$, the bands became fragmented in wide slowly fluctuating domains.
The remaining turbulent patches were seen to decay to laminar flow after long transients when $\RE$ was decreased below $\RE_{\rm g}$.
This bifurcation diagram sketched in Fig.~4, top-right, was confirmed by well resolved, large aspect-ratio numerical simulations in conditions reproducing the experiments (Duguet {\it et al.}, 2010a) or in a narrow oblique domain mimicking a wider geometry (Barkley \& Tuckerman, 2005).

\subsection{Other wall-bounded, spanwise-extended, shear flows\label{S2.3}}
\subsubsection{Circular Couette flow.\label{S2.3.1}}

Couette flow between differentially rotating coaxial cylinders (CCF) is the closest parent to PCF.
It directly depends on two control parameters: the inner and outer rotations rates, $\Omega_{\rm i}$ and $\Omega_{\rm o}$.
Geometric parameters, the inner and outer radii, $r_{\rm i}$ and $r_{\rm o}$, and the length of the cylinders $L$ also play a role, through the radius ratio $\eta=r_{\rm i}/r_{\rm o}$, and the azimuthal and axial aspect ratios $\Gamma_\theta=2\pi r_{\rm m}/d$ and $\Gamma_z=L/d$, where $d=r_{\rm o}-r_{\rm i}$ is the gap and $r_{\rm m}= \frac12 (r_{\rm i} +r_{\rm o})$ the mean radius.
The richness of behaviors displayed by CCF is often depicted in the parameter plane $(R_{\rm o},R_{\rm i})$, where the Reynolds numbers are constructed using the linear velocities $U_{\rm o,i}=\Omega_{\rm o,i}r_{\rm o,i}$, and the gap $d$ as relevant scales, which may not be the best if the plane shear limit $\eta\to1$ is of interest.

The studies  of CCF by Rayleigh (1916) and later Taylor (1923) are landmarks in the theory of flow stability; see (Chandrasekhar, 1961).
From Rayleigh's criterion, flows with co-rotating cylinders may be expected to behave differently than with counter-rotating cylinders.
In the {\it co-rotating\/} case with the inner cylinder rotating faster than the outer one ($0\le \Omega _{\rm o} < \Omega_{\rm i}$),  the flow is always centrifugally unstable, whereas viscous dissipation plays its expected mitigating role.
Accordingly, the first transition is laminar-laminar and supercritical toward Taylor vortices, as a starting point for a globally supercritical cascade to turbulence (Di Prima \& Swinney, 1985).
When the cylinders are {\it counter-rotating\/} faster and faster, the centrifugal instability stay confined in a thinner and thinner layer close to the inner cylinder.
It is soon preempted by shear-driven processes like in PCF.
A cartography of different steady-state flow regimes in the $(R_{\rm o},R_{\rm i})$ plane has been performed by Andereck {\it et al.} (1986) for radius ratio $\eta=0.883$.
Particular regimes of interest to us are the `intermittent-turbulent-spot' and `spiral-turbulent' regimes that destroy the still laminar `spiral'  and `interpenetrating spiral' regimes as $R_{\rm i}>0$ is increased at fixed $R_{\rm o}<0$.
The helical turbulent band regime --  Feynman's celebrated `barber pole turbulence' (Feynman {\it et al.}, 1983) -- was described much earlier by Coles and Van Atta (1962--1967) for a similar value of $\eta$.
As the shear is further increased, the turbulent  helix vanishes leaving room to `featureless turbulence'.
In these early experiments, all with $\eta < 0.9$, the azimuthal aspect ratio was too small to observe more than one helical band.
In experiments with $\eta$ approaching 1 more closely, hence larger $\Gamma_\theta$, Prigent (2001) observed several intertwined helices and was able to measure azimuthal and axial wavelengths of the so-formed laminar-turbulent pattern (Fig.~5, left).
The charts situating the existence domains of the different flow regimes of CCF in the $(R_{\rm o},R_{\rm i})$ plane, with $\eta=0.883$ (Andereck {\it et al.}, 1986), $\eta=0.963$ and $\eta=0.983$ (Prigent, 2001; Prigent {\it et al.}, 2003) clearly indicate a continuous trend toward the diagram corresponding to PCF ($\eta = 1$) once a Reynolds number based on the actual shearing rate is defined ($R^{\rm CCF}=|\eta R_{\rm o} - R_{\rm i}|/2(1+\eta)$) and the mean rotation rate subtracted.
Now considering the turbulent-to-laminar transition, as soon as $\eta$ is sufficiently close to 1, there exists a whole region in parameter space with sustained spatiotemporally intermittent  states in the form of scattered turbulent spots. 
A super-exponential increase of turbulent lifetimes similar to what is observed in HPF  has been reported by Borrero-Echeverry {\it et al.} (2010) but in a geometry corresponding to the occurrence of a single strongly confined spot  ($\Gamma_\theta\simeq22$, $\Gamma_z\simeq34$) similar to those of early experiments in PCF by Bottin {\it et al.\/} (1998), which possibly gives some support to criticisms by Hof {\it et al} (2006) relative to these experiments.
\medskip

\begin{figure}[t]
\begin{center}
\begin{minipage}{0.35\textwidth}
\includegraphics[width=\textwidth]{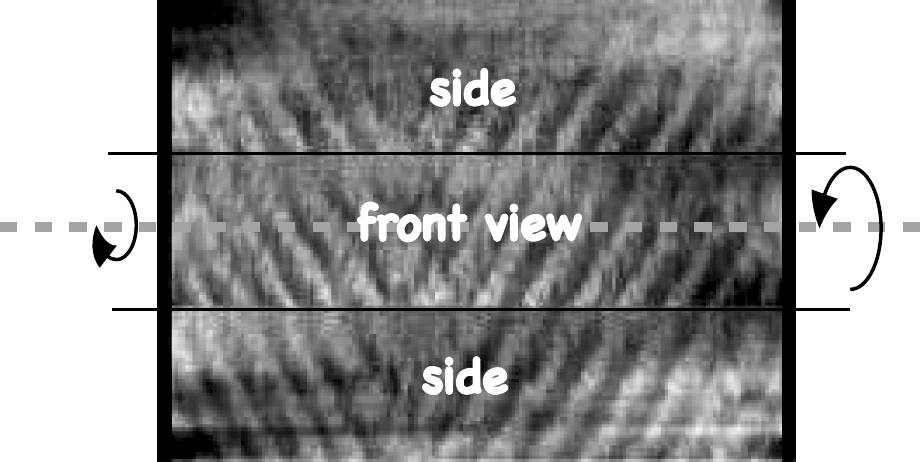}
\end{minipage}\hskip3em
\begin{minipage}{0.45\textwidth}
\includegraphics[width=\textwidth]{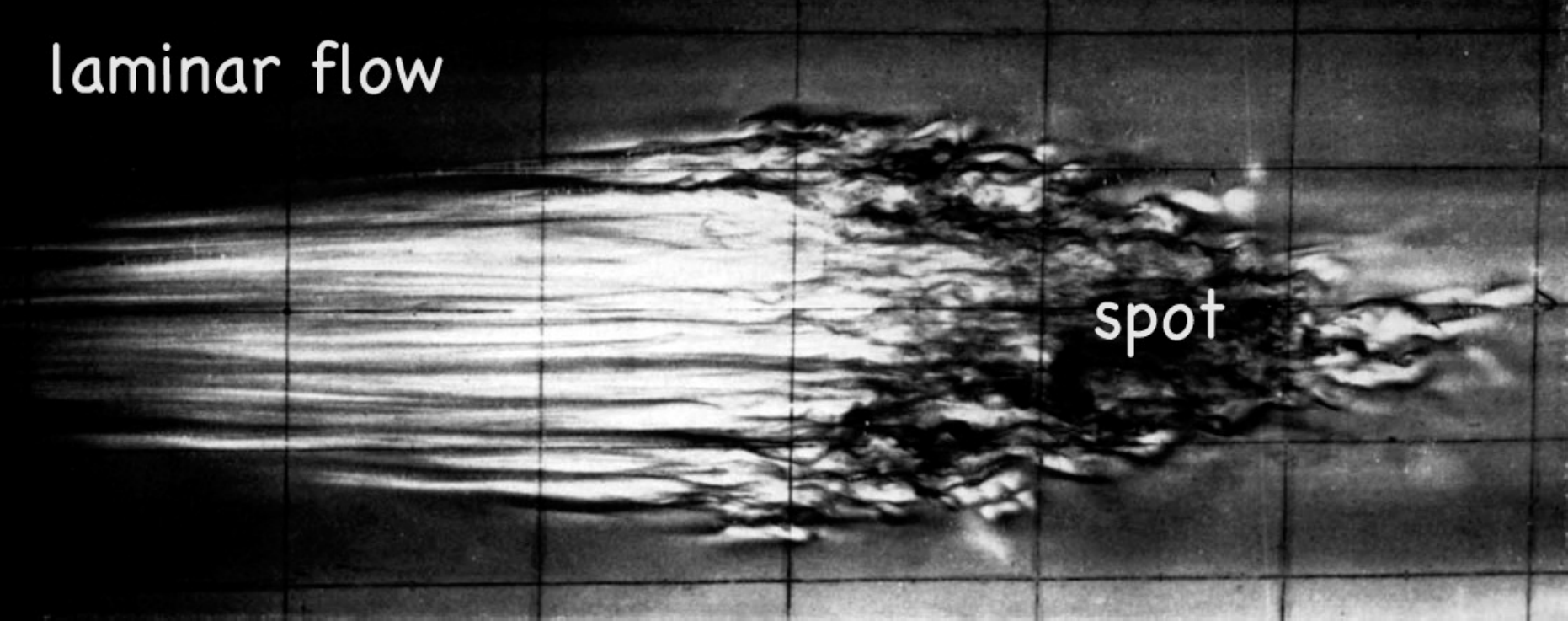}
\end{minipage}
\end{center}
\caption{\label{F5}Left: Spirals in circular Couette flow.  Here, the cylinders' axis direction (gray dashed line), vertical in the experiment, has been shifted to horizontal. The actual diameter of the two cylinders is indicated by the thin black lines; the gap between them is too small to be represented; their contra-rotation is suggested by the arrowed arcs. Mirrors placed behind the cylinders help viewing the spirals from the sides and computerized  image reconstruction gives a $360^{\circ}$ picture of the pattern (Prigent \& Dauchot, 2005). The snapshot shows two spiral domains with different orientations and different pitches transiently composing the pattern. The boundary between them moves so as to eliminate one of the two orientations in the long term. Courtesy: A.~Prigent. Right: Turbulent spot in a boundary layer along a flat plate in a water channel. The perturbation is produced by momentarily poking a rod in the flow downstream of the leading edge of the plate, the spot being photographed further downstreams. After fig.2 in (Edler, 1960).}
\end{figure}

\subsubsection{Torsional Couette flow.\label{S2.3.2}}

In view of its interest for, e.g., turbo-machinery, the flow between rotating discs has also been much studied, see the review by Launder {\it et al.} (2010).
The so-called rotor-stator configuration considered here is with a disk rotating at a small distance of another disk at rest in the laboratory frame (Cros \& Le Gal, 2002).
A comparison with PCF can straightforwardly be drawn when the boundary layers at the disk have merged so that the axial gradient of azimuthal velocity is nearly constant.
This shearing rate however progressively increases from the rotation axis to the outer rim, which allows one to explore the whole range of transitional Reynolds number in a single experiment, implying a difference with previous cases (CCF and PCF).
The flow indeed remains laminar up to some distance of the rotation axis beyond which nontrivial behavior develops as the rotation rate increases.
An instability mode in the form of laminar concentric waves set in at some distance of the center, gets irregular with modulations and phase defects.
Turbulent spirals develop from these defects, roughly comparable to turbulent helices$\mskip1mu/\mskip1mu$bands in the previous cases.
At higher shearing rates, the turbulent spirals leave room to a developed turbulent regime made of a high density of scattered turbulent spots.
Under appropriate rescaling, torsional Couette flow appears remarkably similar to the other flow configurations (Barkley \& Tuckerman, 2007).
\medskip

\subsubsection{Plane Poiseuille channel flow.\label{S2.3.3}}
PPF, the flow  driven by a pressure gradient between parallel plates displays a well-known parabolic profile producing a shearing rate that varies linearly from one to the other plate.
As shown by Orszag (1971), it  is linearly unstable at $\RE_{\rm c}=5772$ ($\RE=U_{\rm cl} h /\nu$, $U_{\rm cl}$: centerline velocity, $2h$: gap between the plates).
In practice, this instability may be bypassed by nonlinear processes when finite amplitude localized perturbations are present.
The possibility of a direct transition to turbulence is therefore open, in the form of turbulent spots and patches.
In early experiments by Carlson {\it et al} (1982), they could be maintained and observed to grow for $\RE\ge 1000$, moving slower than the carrying laminar flow, with head advancing at $\frac23 U_{\rm cl}$ and tail at $\frac13 U_{\rm cl}$, spreading at an angle $\approx 8^\circ$, and possibly splitting further downstream.
The detailed structure of the flow within and around turbulent spots has been recently studied by Lemoult {\it et al.} (2015) to which we refer for an up-to-date bibliography on spots in PPF.
Numerical experiments starting from developed turbulent flow and decreasing $\RE$ have recently shown that turbulence in the form of oblique bands could be maintained at least down to $\RE\approx 800$ in the form of patches and oblique band fragments (Tuckerman {\it et al.}, 2014; Tsukahara \& Ishida, 2014), a fact also recently observed in experiments (Sano \& Tamai, 2015).
When using a Reynolds number based on average shearing rate and half-gap (Barkley \& Tuckerman, 2007), the transition is again seen to take place at values comparable to those for PCF.
\medskip

\subsubsection{Boundary layer flow.\label{S2.3.4}}
 Down along a plate in the absence of applied pressure gradient, the Blasius boundary layer (BBL) flow profile looks close to a hyperbolic tangent.
It is known to be linearly unstable against TS waves at $\RE_{\rm c}\approx520$, $\RE$ being constructed using the local displacement thickness and the speed of the free flow far away from the plate, see  (Schmid \& Henningson, 2001).
A difficulty with this flow is that the thickness of the boundary layer increases with the distance to the leading edge of the plate. This feature can be removed by uniformly sucking through the plate if it is made porous. It is then called the `asymptotic suction boundary layer' or ASBL and  we shall only consider it punctually in the following.

Obtaining a reasonably clean flow is usually an objective of laboratory experiments that can be achieved.
Different scenarios may then be observed depending on the (controlled) level of residual turbulence, involving secondary instabilities of TS waves and their development into $\Lambda$-shaped vortices.
These coherent structures next evolve into {\it spikes\/} before breaking down into turbulent spots that merge to form the developed turbulent boundary layer, as illustrated e.g. by Schlichting (1955).

This scenario  can be viewed as a globally supercritical transition.
However, turbulent spots are better understood as belonging to the nontrivial branch that can be reached as a result of bypassing the regular cascade.
Mechanisms involved in the bypass transition are studied at depth by Schmid \& Henningson (2001), Chap.~9. 
The natural transition path with a high level of residual turbulence is indeed {\it via\/} the nucleation and growth of such turbulent spots, as first shown by Emmons (1951).
 They are arrowhead-shaped and transported by the flow with their heads at about 90\% of the free stream speed and their tails at about 50\%, see Fig.~5 (right).
They expand laterally at an angle of order $10^\circ$ and more slowly in the wall-normal direction ($\approx 1^\circ$). 
The behavior of turbulent spots and the transition to turbulence {\it via\/} turbulent spots has been described many times, e.g. in the review of Riley \& Gad-el-Hak (1985).
\medskip

\subsubsection{Features common to transitional wall-bounded flows from their phenomenology.}

On general grounds, the transition can be said `globally subcritical', with hysteresis upon increasing or decreasing $\RE$.
Transitional Reynolds numbers are large but stay {\it moderate\/}, well-below linear instability thresholds if any (with possibly some reservations for the case of the boundary layer flow).
Furthermore, they appear remarkably similar upon appropriate scale choices based on the local shear (Barkley \& Tuckerman, 2007).

The transition is characterized by the occurrence of localized coherent structures markedly away from the laminar flow profile, with fluctuating but statistically stable or slowly evolving laminar--turbulent interfaces.
Here slowness means time scales much larger than the viscous relaxation time $h^2/\nu$, and thus much much larger than the typical turnover time $U/h$.
By contrast, fluctuations internal to the turbulent domains are fast, with time scales of the order of the turnover time or smaller.

In a large part of the transitional range, and in particular near its lowest end $\RE\gtrapprox \RE_{\rm g}$, localized turbulent structures navigate slightly slower than the local mean speed of the surrounding  laminar flow so that, in a Lagrangian perspective, fluid particles enter the patch smoothly from its rear, get destabilized inside it, follow chaotic trajectories for a while, before leaving it and relaxing towards laminar flow, as illustrated in Fig.~3, 4 (bottom-left), and 5 (right).
This relaxation invariably takes the form of long streaks at the downstream edge of the patch (commonly called the  {\it leading edge\/}) best understood as modulations of the streamwise flow velocity.
On the contrary, the destabilization of the trajectories at the upstream edge ({\it trailing edge\/}) is more abrupt and best visualized from the cross-stream velocity component.
As understood from Fig.~3, this interface can be viewed as the result of a localized Kelvin--Helmholtz-like cascade in the shear layer near the wall at the laminar-to-turbulent entrance of the turbulent domain (Shimizu \& Kida, 2008). 
In PCF, turbulent patches stay more or less at rest in the laboratory frame and leading$\mskip1mu/\mskip1mu$trailing edges have to be appreciated in sublayers symmetrical with respect to the plane at mid-distance between the plates driving the flow in opposite directions.
The laminar-turbulent interfaces then are overhang regions where turbulent flow in a sublayer faces laminar flow in the other.
This configuration was first described by Coles \& Van Atta (1967) in the barber-pole turbulent regime of CCF.

All these features will have to be kept in mind when trying to understand the transition range in detail.
Up to now, `local' and `global' had the meaning implied in the caption of Fig.~1, that is `local' or `global' {\it in phase space\/}, i.e. at a distance infinitesimal or finite from the base flow.
When applied to bifurcation theory, `local' means being amenable to linearization of the equations governing disturbances to the base state and, implicitly, dealing with the finite but small disturbances by some sort of perturbation theory, Fig.~1(a).
In contrast, using the term `global' suggests that finite but large disturbances cannot be teated that way and will ask for a different approach.
This was sufficient to present the phenomenology of the transition as an implementation of the vague but useful idea of globally subcritical scenario, Fig.~1(b2).
In the following, the term `local' will be applied to processes that develop in physical space, as if the flow could be understood as an assembly of `cells' seating these processes (\S\ref{S3}), while the term `global' will relate to the flow system considered as a whole (\S\ref{S4}).
When considering the flow as a whole, we will turn to a statistical account of the transitional regime reminiscent of the thermodynamic approach to macroscopic systems made up of a large number  of particles.


\section{Local viewpoint on the transition\label{S3}}

\begin{figure}[t]
\includegraphics[width=0.33\textwidth]{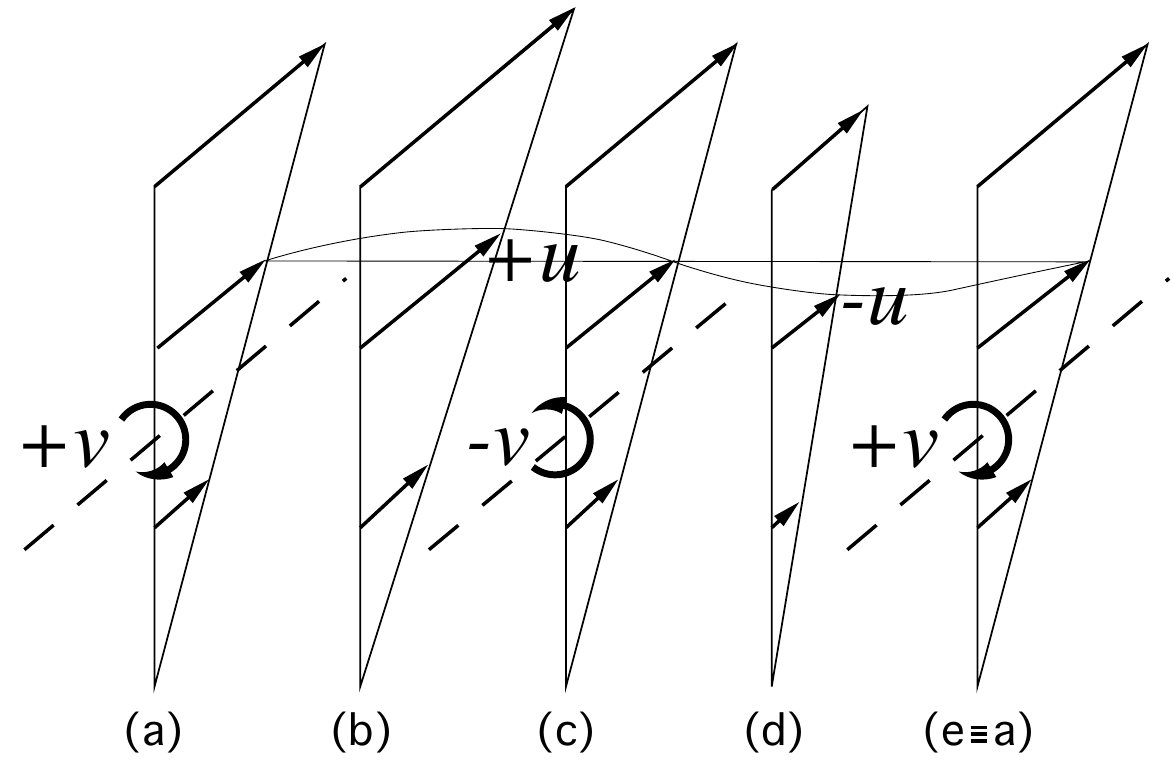}\hfill
\includegraphics[width=0.31\textwidth]{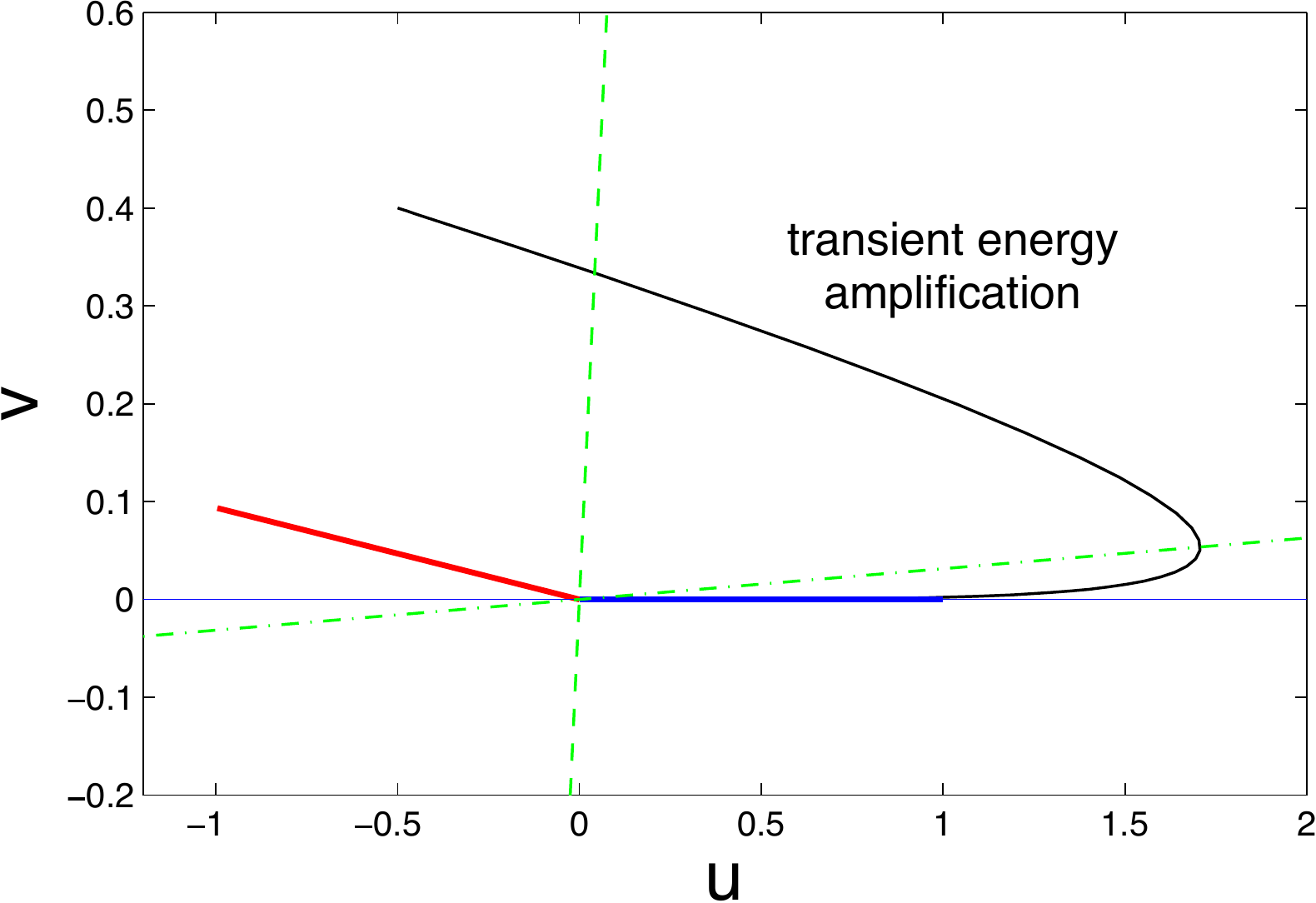}\hfill
\includegraphics[width=0.31\textwidth]{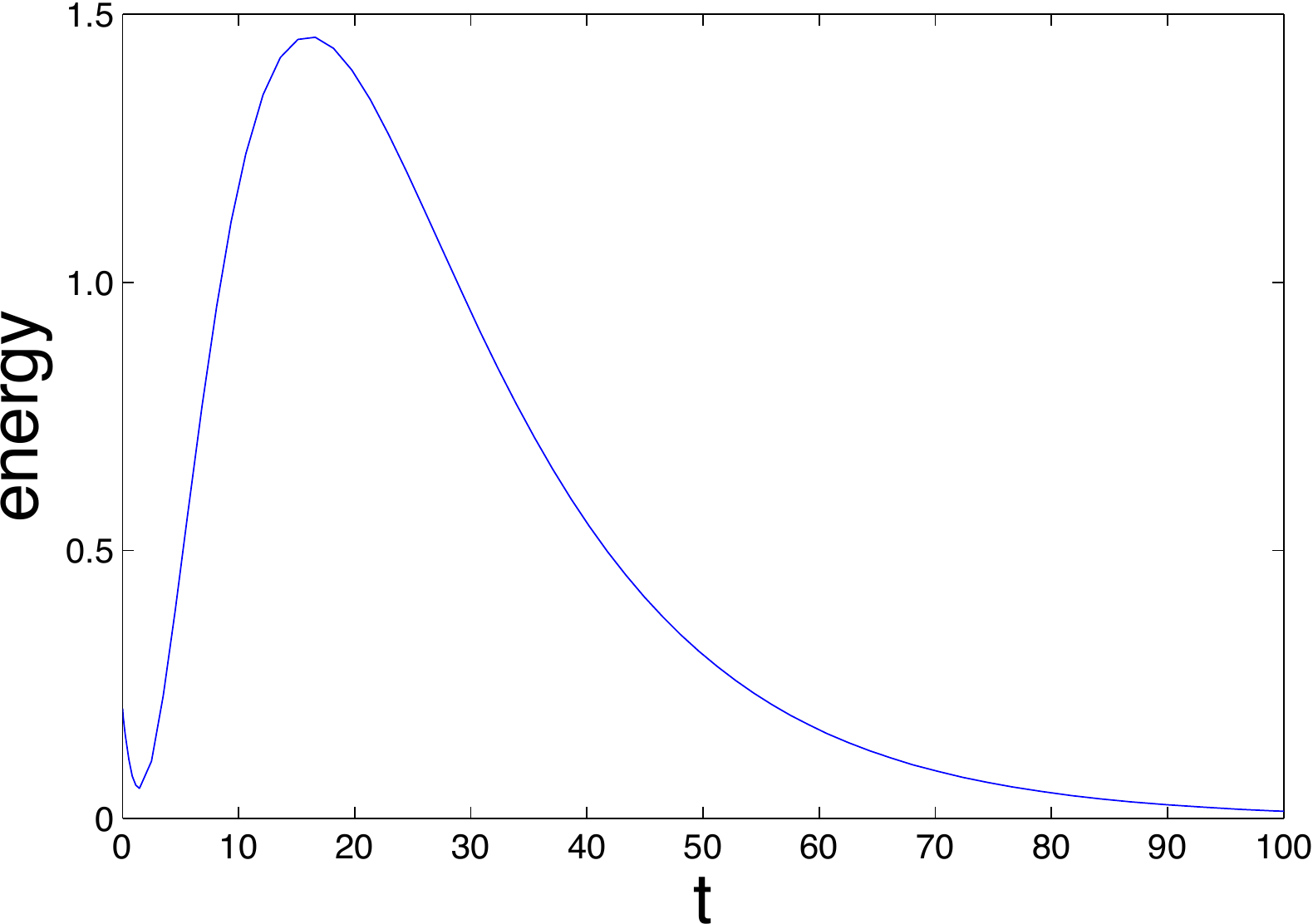}

\caption{\label{F6}Left: In a shear, streamwise vortices produce a spanwise modulation of the streamwise velocity component (streak). Vortices $+v$ at (a) and $-v$ at (c) bring fast fluid from top inducing a high speed streak $+u$ at (b) while vortices $-v$ at (c) and $+v$ and (e) bring slow fluid from bottom inducing a low speed streak $-u$  at (d). Right: Illustration of non-normal linear dynamics using linear system $\dot u = - \alpha_u u + v$ and $\dot v = -\alpha_v v$, with $\alpha_v=4\alpha_u=1/\RE$ and $\RE=8$. In phase space (central panel), $u$'s eigenvector is in blue and the other eigenvector in red; a typical trajectory is depicted in black. Energy $\mathcal E$ is transiently amplified (distance to origin increases) only when the state lies in the sector limited by the two green dashed lines (right panel). Extension to higher dimensions is immediate.}
\end{figure}

Theory discusses the stability of parallel flows in terms of the growth rate of Fourier-analyzed infinitesimal perturbations.
Linear stability predictions then rely on Squire's theorem leading to expect spanwise translationally invariant critical modes at threshold.
In fact this does not mean that, below threshold, when all modes are damped,  the amplitude (energy) of mode superpositions could not grow for a while.
This amplification is due to the non-normal character of  the linear stability operator that admits non-orthonormal eigenvectors and is all the more pronounced that $\RE$ is large (Trefethen {\it et al.}, 1993; Grossmann, 2000), as reviewed by Schmid (2007).
Physically, the underlying mechanism is {\it lift-up\/} (Landahl, 1980) in which a streamwise vortex accelerates (slows down) streamwise flow close to (far from) a wall by mere advection (Fig.~\ref{F6}, left).
A simple model of this important process might be in order.
Calling $u$ the amplitude of the streak and $v$ the strength vortex, from Fig.~\ref{F6} (left) one expects $\dot u \propto v$, where the proportionality coefficient measures the base shear and is set to unity in what follows while the dot indicates differentiation with respect to time.
Modulations are supposed to relax owing to viscous damping so that we can complete the system at the linear stage and write $\dot u = - \alpha_u u + v$ and $\dot v = -\alpha_v v$.
Coefficients $\alpha_u$ and $\alpha_v$ both scale as $1/\RE$ and one may expect $\alpha_u \ll \alpha_v$ from the fact that vortices are more strongly damped than streaks.
A caricature of Navier--Stokes equations would be obtained by completing the linear systems above by nonlinear couplings preserving the kinetic energy $\mathcal E=\frac12(u^2+v^2)$, e.g.
$\dot u = - \alpha_u u + v -uv $ and $\dot v = -\alpha_v v + u^2$ (Dauchot \& Manneville, 1997). 
The interest of the complete model only lies in its full solvability.
Let us here just continue with its linear part (Fig.~\ref{F6}, center and right panels).
The eigenvalues are obviously $-\alpha_u$ and $-\alpha_v$ but the important fact is that the associated normalized eigenvectors $^{\rm t}[1,0]$ and $^{\rm t} [1,(\alpha_u-\alpha_v)]/\sqrt{1+(\alpha_u-\alpha_v)^2}$, respectively (`t' denoting transposition), are not orthogonal for the canonical (energy) scalar product.
As a result, there is a whole sector in phase plane $(u,v)$ where $\mathcal E$ is amplified at the linear stage, namely $v=\rho u$, $\rho\in [\rho_{(-)}, \rho_{(+)}]$, with $\rho_{(\pm)} = (1\pm\sqrt{1-4\alpha_u\alpha_v})/2\alpha_v$. In this simplified case, $\mathcal E$ is essentially the square of the distance to the origin, which is easily extended to the general case. 
The linear operator then operates as a filter letting optimal perturbations emerge and, before decaying, possibly reach a sufficient level to trigger nonlinear dynamics.
This property is an important ingredient of the bypass transition which is best considered through its local facet {\it via\/} the dynamical systems approach (Eckhardt {\it et al.}, 2008).
This framework is indeed the most appropriate to account for the nonlinear dynamics of coherent structures to which turbulent flow can be reduced in the transitional $\RE$ range, \S\ref{S3.1}.
These coherent structures appear to be close to exact solutions of the NS equations under specific periodic boundary conditions,~\S\ref{S3.2}, and, among them, those standing on the boundary of the attraction basin of laminar flow, \S\ref{S3.3}. 
As a down-to-earth output, the theory provides an explanation of the finite lifetime of turbulence close to $\RE_{\rm g}$ in terms of chaotic transients,~\S\ref{S3.4}.

On general grounds, the Navier--Stokes equation can formally be written as a dynamical system $\partial_t X = \mathcal L(\RE) X + \mathcal N\! \mathcal L(X,X)$, with $X\equiv \{\mathbf v(\mathbf x, t),p(\mathbf x,t)\}\in \mathcal X$ infinite dimensional.
Heuristically, dissipation efficiently reduces the dimension of the dynamics by damping out the smallest scales.
This corresponds to the {\it enslaving\/} of the microscopic degrees of freedom to the largest scales of motion directly responding to the extrinsic forcing and the intrinsic instability mechanisms (Haken, 1983).
Reduction to a finite dimensional {\it inertial manifold\/} ensues (Temam, 1990),  that can be viewed as a collapsed phase space on which some {\it effective\/} dynamics develops.
In wall-bounded flows, the relatively moderate value of $\RE$ all along the transitional range is the ultimate justification for a low dimensional effective dynamics. 
Its limitations will however appear when trying to analyze the dynamics of the laminar--turbulent coexistence in extended systems for which {\it spatiotemporal\/} aspects can no longer be ignored (\S\ref{S4}).


\subsection{The minimal flow unit and the self-sustainment of turbulence\label{S3.1}}

Performing numerical simulations in channel flow with streamwise and spanwise periodic boundary conditions  at small distances of the order of the gap between the plates, Jim\'enez \& Moin (1991) explicitly intended to study the mechanics of wall-bounded flow turbulence and associated control strategies.
Such domains were defined so as to carry sustained structures, while nontrivial behavior always decayed in smaller domains at sufficiently large given $\RE$.
The so-introduced {\it Minimal Flow Units\/} (MFUs) were envisioned as building blocks modeling the flow in its transitional range.
This approach was taken over by Hamilton, Kim \& Waleffe (1995) to analyze the mechanism by which nontrivial states can be maintained in the flow, considering PCF in a MFU with size $1.75\pi\times2\times1.2\pi$ in units of the half-gap $h$ between the driving plates.
\begin{figure}[t]
\includegraphics[width=\textwidth]{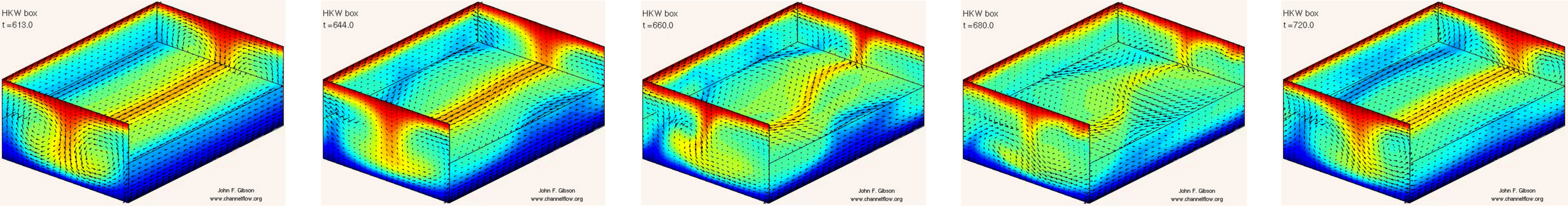}
\caption{Illustration of the different steps of the SSP in a MFU. From left to right: $t=613$, steamwise vortices at start;  $t=644$, intensification of the streaks by lift-up; $t=660$ and $t= 680$, confusing stage of instability and breakdown of the streamwise vortices and streaks; $t=720$, streamwise vortices anew. Color coding of the streamwise velocity component: red: $u_x>0$, blue $u_x<0$. Snapshots taken from videos by Gibson, see {\sc ChannelFlow} WEB site (Gibson, 2012).}
\end{figure}
From careful observations of the flow they identified three steps that are most easily understood by assuming first {\it streamwise vortices\/} occupying the full gap.
These vortices induce {\it streaks\/} by lift-up, as illustrated in Fig.~6 (left).
The streaks correspond to the streamwise velocity component being alternately accelerated and decelerated with respect to the base flow in the spanwise direction.
The so-perturbed base flow then displays inflection points that renders it mechanically unstable.
The streaks then break down, and after some complicated evolution difficult to analyse (Waleffe, 1997), the streamwise vortices are eventually regenerated.
This cyclic regeneration mechanism has been called the {\it Self-Sustaining Process\/} (SSP).
Visual illustrations can be found in (Gibson, 2012), see Fig.~7.
Waleffe (1997) gave an explicit model of it in terms of a truncated Galerkin expansion of the NS equation involving four variables representing the amplitudes of a pair of streamwise vortices, the induced streaks, a streak instability mode, and a mean-flow correction, with quadratic interactions arising from the advection term in the primitive equation.

At this stage it should be stressed that periodic boundary conditions imposed at small distances in the wall-parallel direction(s) play a {\it confinement\/} role.
Combined with dissipation effects, confinement happens to be potent in reducing the number of degrees of freedom, as can be inferred from the obvious coherence and smoothness of flow patterns in Fig.~7, which in turn  supports the low-dimensional dynamical systems approach to the transition to turbulence, like for closed flow systems (Manneville, 1990).
The implicit assumption is that {\it deterministic chaos\/} is the appropriate framework to understand the nature of turbulence (Landau, 1944; Hopf, 1948; Ruelle \& Takens, 1971).

\subsection{Special solutions to the NS equations\label{S3.2}}

In the transitional range, upon increasing $\RE$, nontrivial solutions typically emerge as pairs of states through {\it saddle-node\/} bifurcations at some distance of the base state.
In the pair, one is stable and the other unstable.
The situation is then like in Fig.~1(a2) except that the unstable state belonging to what is called the {\it lower branch\/} is usually not directly connected to the base state.
The strong subcriticality of the problem makes any perturbation approach inadequate and since the direct search for nontrivial solutions far away from laminar flow is not possible, several strategies have been developed to find special solutions to the NS equation.
Analytically the most satisfactory approach is by starting from a known solution to a different problem that can be deformed up to recover the problem of interest, the deformation path being parameterized by some continuous interpolating variable.
This {\it homotopy\/} strategy has been followed by Nagata (1990) who started from Taylor vortices in CCF to obtain a steady solution to the PCF problem by changing the curvature of the base flow.
In a similar spirit, Clever \& Busse (1992) started from Rayleigh--B\'enard cells, adding a simple shear while removing the temperature gradient to reach the PCF case, obtaining the same solution.
Still another strategy was followed by Waleffe (2003) who started from PCF with stress-free boundary conditions and a body force, progressively turning on no-slip boundary conditions while removing the body force.
The bifurcation diagram of such steady  states has been studied, e.g. by Schmiegel (1999) who illustrated the proliferation of steady but unstable solutions in PCF as $\RE$ increases. 
This approach was also adapted to other flows, e.g. HPF in which Faisst \& Eckhardt (2003) and Wedin \& Kerswell (2004) obtained the first traveling waves carried downstreams by the mean flow.

The first periodic solutions in PCF were not obtained in that way but by a clever use of numerical simulations in MFU-sized domains.
Generically, chaotic solutions permanently approach and escape unstable periodic orbits of various periods and shapes.
From a long numerical simulation, Kawahara \& Kida (2001) extracted a periodic-looking sequence and by an appropriate Newton root-finding approach could obtain the first exact periodic solution.
In a similar way, Toh \& Itano (2003) obtained a periodic trajectory in PPF by shooting in appropriate directions of phase space.
In most of these solutions the three steps of the SSP are easily recognized.
Their significance is analyzed by Kawahara {\it et al.} (2012). 
Such solutions could be fugitively spotted in a pipe flow experiment by Hof {\it et al.} (2004).

Studies of exact solutions have been numerous, pushing the approach to a high degree of refinement (Viswanath, 2007).
For example, a detailed cartography of the phase space at MFU scale has been obtained by Gibson {\it et al.} (2008) for PCF.
All these studies are in line with the conjecture by Hopf (as reported by Cvitanovi\'c {\it et al.}, 2013, \S{A5}) that turbulence can be understood as a wandering of trajectories guided by stable and unstable manifolds of unstable limit cycles that proliferate when the control parameter increases sufficiently.


\subsection{The laminar$\mskip1mu/\mskip1mu$turbulent boundary in state space\label{S3.3}}

An important problem relevant to the laminar-to-turbulent transition is to bound the attraction basin of the laminar state by finding the `most dangerous' limit set with a codimension-one stable manifold (a hyper-surface in phase space), and a one-dimensional unstable manifold with neighboring trajectories pointing into the basin of attraction of the base state on one side, and to the hopefully sustained turbulent state on the other side.
Such limit sets, representative of the lower-branch introduced earlier and called {\it edge states\/}, should exist and their defining properties above directly suggests the way to obtain them by dichotomy (Skufca {\it et al.}, 2006).
Within the MFU paradigm they have been obtained in a number of cases, e.g. in HPF (Schneider {\it et al.}, 2007), in PCF (Schneider {\it et al.}, 2008;\dots), in ASBL (Kreilos {\it et al.}, 2013).
Interesting spanwise localized edge states have also been found in streamwise-confined--spanwise-extended geometry, in  PCF (Schneider {\it et al.\/}, 2010a) or ASBL with nontrivial space-time dependence (Khapko {\it et al\/}, 2013).
Even more interesting are the localized edge states obtained in unconstrained geometry, either for HPF (Mellibovsky {\it et al.}, 2009) or for PCF (Duguet {\it et al.}, 2009).
In BBL, the space development of the flow yields growing lambda-shapes vortices with self-similar structure (Duguet {\it et al.}, 2012).

In physical space, these edge states are localized objects immersed in laminar flow with a streaky aspect and mild chaotic dynamics.  
Such structures can be spotted at the end of the turbulent-to-laminar transition in HPF (de Lozar {\it et al.}, 2012) or at the early stage of experiments starting from a high level of random noise in PCF (Duguet {\it et al.\/}, 2010a; see Fig. 2a), which is not surprising in view of the strongly attracting nature of their stable manifold.
Paradoxically, their observation in laminar-to-turbulent experiments is more problematic because this requires the production of low level perturbations of the right shape and amplitude.
The determination of such minimal perturbations can be performed by extending the theory of linear optimal perturbations (that ultimately decay) to the nonlinear domain (Pringle {\it et al.}, 2012; Duguet {\it et al.}, 2013; Cherubini {\it et al.} 2015).
Connection with the standard bypass-transition scenarios (Schmid \& Henningson, 2001) can generally then be made quite easily at the level of the MFU.

\subsection{Transient chaos\label{S3.4}}

Another important use of dynamical systems theory to the transition problem is to help us in interpreting the transient character of turbulence in the near vicinity of the global stability threshold (Eckhardt {\it et al.}, 2008; Eckhardt \& Faisst, 2005).
The proliferation of exact but unstable solutions of all sorts, steady states, limit cycles, etc. with their intricate network of stable and unstable manifolds (homoclinic and heteroclinic tangles) channeling trajectories in state space can be advocated to explain the decay of turbulent states understood as {\it chaotic transients\/} (Lai \& T\'el, 2011).
Exponentially decreasing distributions of transient durations at given $\RE$ are immediate consequences of the presence of a {\it chaotic repeller\/}.
Uniformly seeding the phase space with initial conditions in the region occupied by a tangle indeed yields in a Poisson distribution for the transient durations.
Simple models of crises (Ott, 2006) describe the transition from a repeller into an attractor, leading to predictions for the variation of the mean decay time with the distance to the putative crisis Reynolds number.
These predictions do not seem supported by experiments in HPF (Hof {\it et al.}, 2008) or numerical simulations within the MFU framework for PCF (Schneider {\it et al.}, 2010b) that rather conclude for indefinite super-exponential growth.
Solely focusing on the decay problem, this observation has led to the bold claim that turbulence was always transient, a conclusion somewhat contradictory with our intuition that it should be sustained at least for $\RE$ large enough.
Attacking this conceptual problem is the subject of the next section.
    
\section{Global viewpoint on the transition\label{S4}}

Beyond the advances discussed above, turbulence onset$\mskip1mu/\mskip1mu$decay in conditions of experimental interest still remains an open problem that we must now face.
It is best situated in the context of Fig.~1(b).
Owing to the linear stability in the base flow in the transitional range, it seems clear that we must put some flesh around the globally subcritical skeleton of Fig. 1(b2).
The most recent period has seen interesting advances relative to the nature of the fluctuating states on the nontrivial branch and the sense to be given to the sustained {\it vs.} transient transition as $\RE$ is lowered below  $\RE_{\rm g}$.
In Fig.~1(b), a distance $\Delta$ to the base state was implicitly defined without more specification, and it may seem quite natural to sketch the transition as discontinuous at $\RE_{\rm g}$ using this quantity, following the suggestion by Bottin {\it et al.} (1998) based on early experiments in domains of intermediate aspect ratios.
Whether the transition can be seen continuous when using more suitable quantity, and whether this new quantity grows as a power law with the distance to $\RE_{\rm g}$, is the crux of the problem.
Joint to the observation that we are now considering extended systems with sizes several orders of magnitude larger than the MFU, this property situates the transition problem within the field of {\it critical phenomena\/} familiar in the statistical physics of many-body systems (Kadanoff {\it et al.}, 1967; Stanley, 1999).
Recent claims for {\it universality\/} (Lemoult {\it et al}, 2016; Sano \& Tamai, 2016) have to be understood in this framework.

Before discussing these claims, let us rewind the movie a little since our current understanding strongly depends on previous perspectives taken. 
In the 1970's, mostly following Ruelle \& Takens (1971), the important question of the {\it nature of turbulence\/} was understood as the result of sensitivity to initial conditions inherent in deterministic chaos, the typical dynamics resulting from a few successive bifurcations.
In fact this answer rather relates to the {\it nature of the transition to turbulence\/}, more specifically understood in globally supercritical terms, Fig.~1(b1).
The key word `universality' was inscribed in relevant {\it normal forms\/} controlling a few specific transition {\it scenarios\/}, the well-known sub-harmonic, quasi-periodic, and intermittency routes.  
This was proven appropriate for many flow systems in closed configurations, viz. convection, driven not too far from equilibrium, with dissipation strong enough to maintain dissipative structures coherent at the size of the enclosure.
In confined systems, with all their dimensions of order the wavelength of the instability mode (small aspect-ratio), the mathematical account in terms of ordinary differential equations (ODEs) governing a small number of well-identified mode amplitudes, is indeed suitable.
See related discussions in (Manneville, 1990).

In large aspect-ratio systems, when the system can accommodate many wavelengths, the interest shifts to the properties of {\it patterns\/}, and a corresponding quest for universality of their behavior. 
The common practice is through the transformation of ODE systems for scalar amplitudes into partial differential equations (PDEs) governing amplitude fields.
These spatiotemporal extensions of the amplitudes describe modulations brought to the local variables by topological defects in the patterns, lateral confinement effects, or secondary instability modes.
For a general overview of pattern-formation issues, consult  (Manneville, 1990; Cross \& Hohenberg, 1993; Cross \& Greenside, 2009).
Typically, low-dimensional normal forms (ODEs) of Stuart--Landau--Hopf type are then generalized into PDEs of Ginzburg--Landau type.
Though the generalization is often only based on the problem's symmetries completed by a few phenomenological considerations, the derivation of these PDEs is in principle feasible by means of multi-scale expansions assuming that the modulations to be accounted for are {\it slow in space and time\/}.
The approach is most appropriate in the globally supercritical case, Fig.~1(b1), and key concepts relate to translational invariance and associated phase variables.
An important step in this scenario is the development of phase instabilities.
In particular, from the complex Ginzburg--Landau equation in the neighborhood of the Benjamin--Feir modulational instability threshold, one can derive the  Kuramoto--Sivashisky equation that governs  {\it phase turbulence\/}. Consult (Manneville, 1990; Chat\'e \& Manneville, 1994a) for introductions.
In this context, an essential feature of spatiotemporal chaos is its extensivity, i.e. loosely speaking, the property that the amount of chaos grows linearly with the aspect-ratio since the system can be more or less decomposed into elementary cells that individually contribute to global chaos.

This property is an encouragement to extend the usage of conventional thermodynamic concepts introduced for many-body systems at equilibrium to the transition to turbulence {\it via\/} spatiotemporal chaos.
An important caveat however relates to size effects and whether the so-called {\it thermodynamic limit\/} (infinitely large systems studied at steady-state reached in the infinitely long time limit) is approached sufficiently closely.
While this demand may not be particularly difficult to meet in current thermodynamic systems that scale with the Loschmidt--Avogadro number, the issue is more risky in systems where the `molecular size' is already macroscopic, be it a B\'enard cell or a vortex arrangement in a MFU.
 
Anyway, let us continue with thermodynamics and observe that changes take place through {\it phase transitions\/} of different orders.
When thermal fluctuations are neglected, i.e. phase transitions studied in the mean-field approximation (Kadanoff {\it et al.}, 1967), one can map them onto elementary bifurcations of the kind depicted in Fig.~1 (a).
When fluctuations are taken into account, first-order transitions are characterized by phase coexistence and still remain discontinuous but phase changes are now controlled by large deviations associated to activation processes causing the nucleation of {\it germs\/}.
In contrast second-order phase transitions are characterized by the continuous variation of an {\it order parameter\/} growing from zero in the disordered phase to non-zero in the ordered phase.
In the example of magnetic systems, the magnetization $M$ is the order parameter and varies as $M\propto (T_{\rm c}-T) ^\beta$, where $T_{\rm c}$ is the (Curie) transition temperature.
Exponent $\beta$ is the most immediate example of  {\it critical exponents}.
Within the mean-field approximation one finds $\beta=1/2$ (Kadanoff {\it et al.}, 1967) but thermal fluctuations nontrivially change this value (Stanley, 1999).
`Critical behavior' is essentially associated to scale-free dependence of correlations exactly in the critical conditions.
It manifests itself through a general {\it critical slowing down\/} of the spatiotemporal dynamics in their neighborhood.
A~set of exponents then characterizes the approach to criticality, controlling the divergence of appropriately defined correlation lengths and relaxation times.
In turn this divergence renders of the precise nature of the microscopic interactions less relevant near the critical point, which supports a specific concept of {\it universality\/}, namely that all systems with the same qualitative properties (symmetry of the order parameter, dimension of the physical space) will be quantitatively characterized by the same set of exponents.

As to the transition to turbulence in wall-bounded flows, the understanding of its nature, especially regarding its universality is presently a hot topic.
Is `universality' that of ({\it i\/}) dynamical systems though normal forms and bifurcation scenarios, of ({\it ii\/}) pattern formation, its envelope and phase or defect dynamics, or of ({\it iii\/}) critical phenomena, with their scale free behavior at the thermodynamic limit?
The last possibility, most relevant to subcritically transiting  extended systems,  is scrutinized in what follows.

Up to now, most of the local information on transitional processes has been obtained in the MFU-restricted, low-dimensional dynamical system framework, especially about the upper-branch states away from laminar flow and the SSP mechanism on the one hand, and about the lower branch edge states sitting on the basin boundary of the base state on the other hand.
The effective confinement inherent in this approach is unable to account for space modulations observed in actual experiments that are all performed in systems much wider than the dimension relevant to the wall-normal direction, pipe radius or gap between plates.
Whereas, in the transitional range, coherence is maintained along that direction, large scale dependence along the one (tube axis) or two (in-plane) complementary directions has to be dealt with.
Main difficulties arise from the wild and complex time dependence of the chaotic upper-branch states and the  short-range space dependence with abrupt laminar-turbulent interfaces allowed by the subcritical character of these states.

Deductive theoretical approaches directly from the NS equations are hopeless and, in an attempt to make some progress, one has to turn to modeling with much phenomenological input from empirical observations.
Options more radical than extensions of the Ginzburg--Landau formalism have to be taken.
For example, staying within the pattern universality perspective, a quintic nonlinear extension necessary for subcriticality only generates pulses or fronts with regular dynamics (van Saarloos \& Hohenberg, 1992), possibly quite complicated (Bottin \& Lega, 1998), but reveals itself insufficient for stochastically-evolving puff$\mskip1mu/\mskip1mu$spots boundaries or the fluctuating turbulent bands.
In order to tackle this issue, Pomeau  (1986, 1998, 2015) advocated the study of the subcritical transition to turbulence in terms of {\it directed percolation\/} (DP) accounting for the competition between a spatiotemporally active regime and a globally everywhere-inactive {\it absorbing state\/}  (Henkel {\it et al.}, 2008).
In subsection~\ref{S4.2} we give a very brief overview of this prototype of non-equilibrium phase transition and its critical properties (DP universality class).
DP is a fully stochastic process defined on a lattice that has shown its relevance for more general determinisitic systems displaying {\it spatiotemporal intermittency\/} (STI), as briefly reviewed by Chat\'e \& Manneville (1994b). 
STI is conveniently modeled using lattices of coupled maps that straightforwardly implement the DP problematics in a deterministic context and can be studied using the same tools.
Subsection~\ref{S4.3} is devoted to the examination of the extent to which  the transition to turbulence fits the STI framework and may belong to the DP universality class.

\subsection{Directed percolation: critical properties\label{S4.2}}

As already mentioned, coexistence in phase space associated to sub-criticality translates itself into coexistence in physical space of domains homogeneously filled with one or the other of different states.
When at least one of the competing states is chaotic, the position of the domain boundaries can be expected to fluctuate, due to the intrinsic amplification of small perturbations  away from strict uniformity of the local state arrangement.
Pomeau's idea was that the fluctuating dynamics of fronts between laminar and turbulent domains could be modeled by DP.
This purely stochastic process is also invoked to describe such phenomena as forced flow through porous media, forest fires, or epidemics.
It can be defined as a two-state probabilistic cellular automaton evolving on a $D$-dimensional lattice (Henkel {\it et al.}, 2008).
Each cell can be in one out of two possible states, {\it active\/} or {\it inactive\/}.
The state of a cell at a future time $t+1$, depends on the states of neighboring cells at time $t$ {\it via\/} the random drawing of transition probabilities, with the rule that a cell cannot become active by itself or if all its neighbors are inactive, but only by {\it contamination\/} when at least one neighbor is active.
When the contamination probability is too low, the infection stops and the whole system is trapped into an {\it absorbing state\/} where all cells are inactive.
The system can thus be envisioned statically on a $(D+1)$-dimensional lattice with bonds oriented in the supplementary dimension representing time, hence the `directed' qualifier to the term `percolation' that refers to our interest in clusters of activity.

Assuming that the process is controlled by a single probability $p$, a transition is observed between a phase with sustained activity for $p> p_{\rm c}$ and the absorbing phase.
The order parameter of this transition is the mean fraction of active states $\rho$ of the system (mean activity), and the reference system described above displays a continuous transition characterized by a set of critical exponents.
The mean activity grows as $\rho\propto(p-p_{\rm c})^\beta$, while near threshold space correlations decrease as $\exp(-x/\xi_\perp)$, $\perp$ featuring the $D$ physical space dimensions, and time correlations as $\exp(-t/\xi_\parallel)$, $\parallel$ referring to the time direction. 
As alluded to previously, the essential feature is the divergence of the coherence lengths as the threshold is approached, that is: $\xi_{\perp,\,\parallel} \propto (p-p_{\rm c})^{-\nu_{\perp,\,\parallel}}$ ($\nu_{\perp,\,\parallel}>0$), yielding a scale-free behavior at threshold.
DP defines a {\it universality class}, i.e. a class of processes characterized by the same set of critical exponents.
Qualitative equivalence implied by universality is conjectured to rely on ({\it i\/}) the existence of a single absorbing state, ({\it ii\/}) the continuous variation of the order parameter through the threshold, ({\it iii\/}) short range interactions, and ({\it iv\/}) no hidden symmetries nor quenched disorder, in witness whereof all systems with these characteristics will share the same set of exponents.
There are no known exact results but exponents can be determined from numerical simulations.
For $D=1$ one gets $\beta\approx 0.28$, $\nu_\perp\approx 1.1$, $\nu_\parallel\approx 1.7$, and for $D=2$,   $\beta\approx 0.58$, $\nu_\perp\approx 0.73$,  $\nu_\parallel\approx 1.3$. For more information consult the reference book by Henkel {\it et al.} (2008) and for $D>2$, of less direct interest here, the recent work by Wang {\it et al.} (2013).

\subsection{The directed percolation issue in spatiotemporally-intermittent transitional wall-bounded flows\label{S4.3}}

Obviously, the `turbulent' and `laminar' local flow regimes respectively correspond to the `active' and `inactive' states.
The absorbing character of the inactive state is a byproduct of the subcritical character of the transition since laminar flow can only be destabilized by finite-amplitude (turbulent) perturbations.
As noticed previously, local chaos in the active state brings in the necessary stochasticity.
The transient nature of the local (MFU-scale) chaos when $\RE$ is not large enough is responsible for global decay but now with a definite spatiotemporal flavor, owing to the contamination property.
Accordingly, the onset of sustained turbulence therefore results from the conversion of {\it local transient temporal\/} chaos into {\it sustained global spatiotemporal\/} chaos when $\RE$ is large enough, whereas the supposed universality exempts us to discuss the very nature of decay and contamination processes provided they are sufficiently short-ranged.
Despite the possibly transient character of the local turbulent state even for much larger values of $\RE$ (Hof {\it et al.}, 2008; Schneider {\it et al.}, 2008), it is the topological nature of the DP transition -- with decaying active clusters below threshold turning into invading ones above -- that comes and closes the controversy about the finite-lifetime character of turbulence put forward by Hof {\it et al.} (2006), agreeing with our common intuition that, for sufficiently large $\RE$ turbulence is sustained.

While the relevance of DP to transitional wall-bounded flows is evident at a formal level, the universality question remains debatable, owing to the fully stochastic framework, appropriate or not, used to describe phenomena governed by some underlying deterministic dynamics with possibly long-range effects, here the NS equation.
As a matter of fact, before being tested on transitional wall-bounded flows for which it was designed, Pomeau's conjecture was scrutinized in the general context of {\it spatiotemporal intermittency\/} where the stochastic framework of probabilistic Boolean automata lattices is replaced by deterministic iterations over continuous variables coupled by diffusion.
Such {\it coupled map lattices\/} (CMLs) offer a convenient intermediate modeling level between the NS equation level and the DP scheme, by avoiding to turn to random systems too early.
The situation described above for DP is indeed recovered in a deterministic context by taking minimal local iterations displaying transient chaos on a repeller (active state) in competition with a fixed point (inactive state), as described in (Chat\'e \& Manneville, 1994b).
Numerical simulations however showed that, though the conditions to observe DP-universality were apparently fulfilled, critical exponents could indeed be estimated that were close to but not exactly those of the DP class in systems of intermediate size followed during too short durations.
Grassberger \& Schreiber (1991) explained the fact by noticing that, despite randomness implied by transient chaos, determinism leaves its mark by introducing long time correlations {\it via\/} propagating, finite lifetime, localized coherent structures, which forces to redefine elementary space and time scales much larger than the original space and time steps, rendering the {\it thermodynamic limit\/} extremely difficult to achieve.

Beyond the case of abstractly defined CMLs (Chat\'e \& Manneville, 1994b), spatiotemporal intermittency has been the subject of several concrete studies, e.g. in 1D closed flow configurations (Daviaud, 1994), but with little confirmation of the actual relevance of DP universality.
An exception is the turbulent-turbulent transition observed in nematic liquid crystals with two competing types of electro-hydrodynamic modes (Takeuchi {\it et al.}, 2009), where the 2D DP class was clearly identified.
The case of subcritical wall-bounded flows is of more recent interest.
A interesting attempt is to be found in the work of Barkley (2011) who modeled the dynamics of puffs and slugs in HPF by a 1D CML with well-designed transient local dynamics and diffusive spatial coupling closely related to what has just been described.
In this model, the decay {\it vs.} splitting issue seems to fall into the 1D DP class, strongly suggesting the same property for the experimentally observed transition at $\RE_{\rm g}\approx2040$ (\S\ref{S2.1}).
Sill in a basically 1D configuration, Lemoult {\it et al.} (2016) considered cylindrical Couette flow (CCF) in a short cell with small curvature and long perimeter and performed a similarly 1D numerical experiment sampling CCF along an appropriately oriented spiral, as a continuation of the work in (Shi {\it et al.}, 2013).
In both cases they found spatially distributed turbulent patches that display exponents compatible with the one-dimensional DP universality class.
Sano \& Tamai (2016) recently claimed similar adequacy of the 2D DP universality class in their study of channel flow (PPF) at relatively large aspect ratio under turbulent forcing, which however requires support under broader conditions.
At this point, one could also add that the DP-inspired statistical approach discussed above has successfully led Kreilos {\it et al.} (2016) to design an empirical stochastic automata lattice modeling the transitional boundary layer, which might prove both more promising and far-reaching.

When invoking DP and its universality class, it is implied that the transition is continuous, which, from the first principles, has no obvious reason to hold.
In fact, Pomeau (1998) also considered the discontinuous case akin to a first-order thermodynamic phase transition.
In thermodynamics at equilibrium, this last case corresponds to free energies with several minima corresponding to the different coexisting phases separated by energy barriers.
The transition is then controlled {\it large deviations\/} overcoming the barrier, which physically corresponds to the {\it nucleation\/} and subsequent evolution of {\it germs\/} of one phase into the other.
This discontinuous case is less sensitive to size effects than the continuous case since it does not imply the divergence of spatiotemporal correlations but rather the nucleation of finite size germs.
Accordingly, consistent results are expected as soon as the lateral extension is sufficiently larger than the typical size of germs able to trigger the transition.
This extreme-event problem  is standard in the field of stochastic systems displaying several competing phases.
The laminar-to-turbulent transition is not a good candidate since the evolution of turbulent puffs, spots, and the like, is only mildly chaotic. 
In contrast, the nucleation of a laminar trough within the turbulent phase during the turbulent-to-laminar transition indeed presents itself as a large-deviation probabilistic problem.
Simplified analog models of PCF, either in terms of CMLs (Bottin \& Chat\'e, 1998), or partial differential equations (Manneville, 2009), exhibit such first-order features.
Somewhat outside the continuous (2nd order) {\it vs.} discontinuous (1st order) debate, but still in the same context, extreme events have tentatively been invoked to interpret the double exponential behavior of the decay or splitting mean waiting times in HPF (Goldenfeld {\it et al.}, 2010; Duguet, private communication).
Similarly, a study of PCF at moderate aspect ratio showed that the decay of turbulence could be analyzed in terms of extreme value laws (Faranda {\it et al}, 2014).

In fact, slightly under-resolved DNS of PCF near $\RE_{\rm g}$ have shown that the decay and growth of turbulent patches involve a mixture of processes (Manneville, 2011, 2012a).
First, as mentioned above, large deviations are responsible for the opening of wide laminar gaps in continuous bands during decay  for $\RE\lesssim \RE_{\rm g}$  
and large scale reorganizations associated to spot splitting during pattern growth for $\RE\gtrsim \RE_{\rm g}$.
Next, besides these extreme events at the size of band width or spot size, small-scale processes strongly reminiscent of stochastic contamination processes assumed in the DP approach are taking place.
Depending on the value of $\RE$, these small-scale processes show a systematic bias towards decay (regular receding of turbulent band fragments) or growth (preferential oblique expansion).
These features qualify for a first-order phase transition close at the Maxwell plateau rather than for  criticality at a DP-like threshold.
In addition, large-scale secondary flows generated by Reynolds stress inside the turbulent patches, though expected to decay fast away from their sources, strongly influence the small scale stochastic growth$\mskip1mu/\mskip1mu$decay of coherent structures at the laminar-turbulent interfaces and contribute to their reshaping (Duguet \& Schlatter, 2013; Couliou \& Monchaux, 2015; Manneville, 2015b).
In the same way, even in the case of HPF that seems the best candidate for the application of a continuous DP transition, the identification of puffs as elementary coherent structures amenable to local decay and growth by splitting is not evident.
It seems that the individual `cells' participating in the transition process should include long fluid sections where already laminar recirculation flow is still evolving to recover the parabolic base profile (Hof {\it et al.}, 2010).     
So, a quite complicated picture therefore emerges, involving widely different  scales and timings.
While one of the properties characterizing the DP universality class seems beyond questioning, namely the existence of a single absorbing state, two of the other assumptions, the short-range character of couplings and the continuous nature of the transition remain to be ascertained on a case-by-case basis, in connection with the relevance of the thermodynamic-limit caveat.

To wrap up this discussion, appeal to universality in the sense of critical phenomena at a second order phase transition is a stimulating conjecture (Pomeau, 1986).
Indications of the relevance of the DP universality class have been found in 1D cases, in a CML model of HPF (Barkley, 2011) and in numerical simulations and experiments in short-cylinder  CCF (Shi {\it et al}, 2015), and in a 2D large aspect ratio PPF (Sano \& Tamai, 2015).
However, the study of critical properties is notoriously difficult, requiring to approach the long-time, large-system, thermodynamic limit.
Here the relevant coherent units have not molecular space-time scales like in thermodynamics, but already macroscopic dimensions scaling with the typical wall-normal unit length.
These coherent units typically scale as the MFU  {\it a minima\/} but are in practice already much wider.
For example streamwise correlation beyond MFU size, in fact of the order of the width of turbulent bands, has to be taken into account to really understand the temporal-to-spatiotemporal transition in PCF (Philip \& Manneville, 2011) and large scale recirculation around localized structures (Duguet \& Schlatter, 2013; Hof {\it et al.} 2010) likely need to be included in the definition of local automata, stochastic or not, involved in the statistical physics approach.
Accordingly, usual numerical or experimental systems may not contain sufficiently many of them and finite size effects have to be faced.
As to the continuous {\it vs.} discontinuous issue, depending of circumstances, CML models with STI may display DP-like 1st or 2nd order transitions and, when 2nd order, may present scaling behavior outside the DP universality class due to hidden long range correlations of deterministic origin (Grassberger \& Schreiber, 1991).
Though the transition to$\mskip1mu/\mskip1mu$from turbulence in wall-bounded flows is undoubtedly a non-equilibrium transition with an absorbing state (Henkel, {\it et al.}, 2008), its belonging to the strict DP universality class appears to be a question of relatively limited interest to which a definitive answer might be difficult to obtain despite recent evidence.
Like the liquid-gas transition in thermodynamics which is 1st-order in general and 2nd-order only when specific conditions are met, the laminar-turbulent transition has no obvious reasons to be discontinuous rather than continuous with exponents.
Only a comprehensive theory of transitional turbulent flow, like the van der Waals theory of gases, would really explain the conditions under which expecting DP criticality is legitimate.
This should be the objective of realistic modeling to which we now turn.

\subsection{Understanding \textbf{\textit{via\/}} modeling \label{S4.4}}

Statistical-physics approaches invoking e.g. directed percolation or large deviations are just a compact way to describe the transition with little physical interpretation of the processes at stake.
Some progress has been achieved in view of understanding the transition by borrowing ideas developed in the chemistry of reaction-diffusion (RD) systems, a field where propagation  of concentration pulses or steps is a central issue.
In such systems an important distinction has to be made according to whether the system as a whole is just {\it excitable}, that is, returns to its base state after the passing of a flash of perturbation (e.g. nerve impulse), or {\it bistable\/} with the most stable state invading the least stable one (e.g. a flame front between fresh and burnt gases).
This analogy works remarkably well to describe the transitional regime of HPF, with the transformation of puffs in the excitable regime into slugs in the bistable regime as $\RE$ is increased.
A complete model has been derived by Barkley (2011), then improved and fully exploited in (Barkley {\it et al.}, 2015).
It involves just two variables governed by two 1D partial differential equations (or two 1D CMLs in its discrete space version).
The variables characterize the local intensity of turbulence and the speed on the pipe's centerline, respectively, while the coupling terms feature all important processes at work.
Advection is treated in a realistic way and the switching from the excitable regime to the bistable state is controlled by a single parameter mimicking $\RE$.
The detailed scenario of the HPF scenario is then reproduced quantitatively just by tuning one parameter with a few others appropriately adjusted.

In the same vein, Goldenfeld and collaborators (Shih {\it et al.}, 2015) have proposed an interpretation of `activator-inhibitor' reactions as species interactions of `prey-predator' type in ecology and constructed a corresponding (quasi-)1D probabilistic automaton with rules adapted from beforehand numerical simulations of pipe flow in a very short tube.
Appropriate choice of probabilities led to general agreement with experimental findings and equivalence with 1D DP was demonstrated without surprise in view of the way the model was constructed, still somewhat far from a derivation from first principles.

Still in a RD framework, we suggested a very heuristic level that the regular laminar-turbulent pattern in the upper transitional range of PCF could be understood as the result of a Turing instability -- a standard pattern forming process in RD systems -- of the high-$\RE$ featureless turbulent regime upon decreasing $\RE$  (Manneville, 2012b).
An explicit spatiotemporal extension of the model in (Waleffe, 1997) was proposed that accounts for such a patterning as soon as the diffusion of the transverse disturbances measuring the intensity of turbulence is substantially faster than that of the streamwise perturbation, which is physically sound but still far from being justified from just considering the primitive equations.

The nonlinear terms in (Barkley, 2011) or (Barkley {\it et al.}, 2015), as well as the rules in (Shih {\it et al.}, 2015), were chosen according to phenomenological considerations.
In (Manneville, 2012b), nonlinearities were borrowed from (Waleffe, 1997), where they were derived from the NS equation within a MFU assumption {\it via\/} appropriate truncation of a Galerkin expansion.
Getting rid of this assumption and projecting the NS equation on adapted wall-normal bases  yields systems of partial differential equations governing the amplitudes of the corresponding modes (instead of ordinary differential equations in the MFU framework).
By construction, the obtained systems have the same structure as the NS equation but can be truncated at different level to test the role of the wall-normal resolution on the pattern formation.
The numerical simulation of these systems  (Seshasayanan \& Manneville, 2015) has shown that a minimum of wall-normal structure has to be included to obtain the laminar-turbulent patterns typical of transitional PCF, whereas lowest order truncation already accounts for disordered laminar-turbulent coexistence (Lagha \& Manneville, 2007a).
Large scale secondary flows generated by Reynolds stresses within the turbulent patches are realistically rendered by the models whatever the truncation level (Lagha \& Manneville, 2007b). However, up to now their analysis (Manneville, 2015b) has produced nothing as simple as the Turing-like mechanism in (Manneville, 2012b) able to explain the patterning, so that its origin remains mysterious.

Very recently, Chantry {\it et al.\/} (2016) opened an interesting path by pointing out that the transitional regimes of the flows of interest here, PCF, PPF, and HPF, could be mapped one onto the other by appropriately rescaling the velocity and the wall-normal dimension, hence the Reynolds number.
The remark relies on comparisons of the `bulk' properties of these flows (i.e. excluding processes localized in the boundary layers) to those of the Waleffe flow, a shear flow configuration with stress-free boundary conditions and a sinusoidal body-force driving.
Extending an approach initiated in (Manneville \& Locher, 2000), they further derived a simplified Galerkin model reproducing the oblique laminar-turbulent patterning better and at lesser analytical cost than in (Seshasayanan \& Manneville, 2015) who dealt with a shear flow driven by no-slip boundary conditions, but in the same spirit.
Perspectives opened by this work are highlighted in (Manneville, 2016).

These different modeling attempts clearly indicate that we now have appropriate tools to understand laminar-turbulent coexistence in the upper transitional range, and in a second instance, also possibly the statistical nature of processes involved in its decay at its lower end.

\section{A summary and some conclusions, including open issues\label{S5}}

The transition to turbulence in wall-bounded flows is a long standing problem first posed by Reynolds at the end of the XIXth century, see (Eckhardt ed., 2009).
Standard (linear) stability theory proved unable to explain it and empirical observations long remained hard to decipher.
Clarification only partly came from progress in recent experiments that, though technically more elaborated, were conceptually not much different from those of the Founding Father.
In fact, the situation profoundly evolved during the past twenty years due to a change of  perspective fruitfully combining new concepts and tools from mathematics (chaos theory), computer science (large scale simulations, search for special solutions), and  statistical physics (phase transitions), while a single of them would possibly not have led to our present understanding level.
A pregnant question about the nature of turbulence (Landau, 1944) previously dealt with in terms of chaos (Ruelle \& Takens, 1971)  has thus recently been re-focused on spatiotemporal issues and 
new tools$\mskip1mu/\mskip1mu$concepts have been put forward to complete and correct shortcomings of a purely temporal approach.

The transition to turbulence in wall-bounded flows does not follow a progressive, globally super-critical scenario.
On the contrary, it is direct and marked by the coexistence of laminar flow and localized patches within which the flow is strongly distorted and can be considered as turbulent.
An essential feature is that the transition takes place at moderate Reynolds numbers, sufficiently low that, still free from linear instability modes, laminar flow remains a possible stable regime, but high enough that the nonlinearity of the Navier--Stokes equation permits the existence of nontrivial flow regimes in the form of coherent structures. 
Two main problems have then to be faced.
First,  for the laminar-to-turbulent transition, what is the nature of the state inside the patches and how does it emerge?
Second, at the turbulent-to-laminar transition, how do we understand the laminar-turbulent patterning, from both a physical viewpoint (mechanism underlying the alternation) and from a statistical viewpoint (decay of chaotic flow and contamination of laminar flow)?
Laminar or turbulent, these flow regimes clearly do not play symmetric roles.
The amplitude of fluctuations around laminar flow can be considered as small.
In clean experiments the level of residual turbulence is indeed very low but, even in not-so-clean experiments, the transition from laminar has to be triggered by perturbations controlled from the outside.
In contrast, turbulent flow is permanently agitated by large fluctuations around its mean, so large that the laminar bottom line can be reached locally, which gives its distinctive flavor to turbulent decay.

Two examples have been treated in more detail, the flow in a pipe (HPF) and the simple shear flow (PCF), both are linearly stable for all $\RE$, which exacerbates the subcritical character of the transition.
They share similarities that, in turn, are prototypical of what takes place in other flows displaying laminar-turbulent coexistence at moderate $\RE$, with the turbulent state belonging to a branch disconnected from the laminar base state branch and bifurcating in (phase-space) limbo of turbulence.

Since there are no relevant linear instability modes (Romanov, 1973; Salwen {\it et al.}, 1980), the amplitude and shape of localized perturbations able to trigger the transition away from laminar flow are both important.
Transient energy growth {\it via\/} lift-up (Landhal, 1980), a linear process related to the non-normal character of the operator controlling the stability of the base flow (Trefethen {\it et al.}, 1993), appears to be a major ingredient of the self-sustainment of turbulence.
The concept of Minimal Flow Unit (Jim\'enez \& Moin, 1991), the smallest periodic domain supporting nonlinear nontrivial states,  has been instrumental in identifying how this mechanism operates in coherent structures at a local scale (Hamilton {\it et al.}, 1995; Waleffe, 1997).
An important stream of research has then been dedicated to this problem using the tools of dynamical systems theory, culminating ({\it i\/}) in the understanding of turbulent flow in relation to exact solutions to the NS equation (Kawahara {\it et al.}, 2012), especially periodic orbits  and the entanglement of their stable and unstable manifolds (Cvitanovi\'c {\it et al.}, 2013),  ({\it ii\/}) in the analysis of the local breakdown of turbulence as a result of transient chaos associated to ensuing homoclinic tangles in phase space (Eckhardt {\it et al.}, 2008), and  ({\it iii\/}) in the identification of edge states lying on the border of the attraction basin of the laminar base flow (Skufka {\it et al.}, 2006).

Up until recently, the dynamical systems approach has been developed within the MFU context in phase spaces of reduced effective dimensions unable to account for the turbulence intensity modulation  in physical space.
Beyond the strict MFU assumption, the observation of modulated states, in particular localized states, was exceptional (Duguet {\it et al.}, 2009; Schneider {\it et al.}, 2010).
In order to deal with really spatially extended systems from the start, a radically different approach was proposed by Pomeau (1986, 1998, 2015) starting from an analogy between the subcritical transition to turbulence and non-equilibrium phase transitions into absorbing states, and more particularly with {\it directed percolation}, a spatiotemporal, fully {\it stochastic\/}, process defined on a lattice, a thorough presentation of which is given in (Henkel {\it et al.}, 2008).
This approach sets the problem in the framework of {\it critical phenomena\/} in statistical physics and stresses on {\it universality\/} properties of the second-order phase transitions in thermodynamics (Stanley, 1999).
In this context, indicators of criticality (order parameter, correlation lengths) depend on the relative distance to the transition point through power laws with exponents depending only on gross features such as space dimensionality and symmetries but  not on the details of the processes involved.
Signs that the directed-percolation universality class could be relevant have been collected in 1D configurations for HPF (Barkley, 2011) or PCF (Shi {\it et al.}, 2015) and in 2D for PPF (Sano \& Tamai, 2015).
However, the universality issue, while appealing at the qualitative level, is not settled quantitatively, all the more that the continuous character of the transition, mandatory to fit the critical phenomena approach, is not ascertained.
There is indeed also the possibility of a discontinuous first-order transition (Pomeau, 1998) with associated large deviations ending in the nucleation of germs, here finite-size patches of laminar flow within turbulent flows.
At least band fragmentation in the lowest part of the transitional range of PCF in 2D can be interpreted in these terms (Faranda {\it et al.}, 2014).

The transition to turbulence in wall-bounded flow is not only a problem in statistical physics, but first a problem in fluid mechanics.
As such, it raises questions about the feedback between coherent structures at the scale of the typical wall-normal dimension and secondary flows evolving at much larger scales.
Reynolds stresses generated in far-from-laminar domains drive such large scale flows.
Playing a role in turbulent spot growth (Lagha \& Manneville, 2007b; Duguet \& Schlatter, 2013), they demands further scrutiny (Manneville, 2015b).

Still in a similar hydrodynamic context, free shear flows such as shear layers, jets or wakes, experience a transition between two different mixing regimes as $\RE$ is decreased.
This transition is between fine scale {\it mixing\/} in the developed turbulent regime at higher $\RE$ and a {\it stirring\/} regime with larger chaotic coherent structures at lower $\RE$ as reviewed e.g. by Dimotakis (2000).
It may be attributed to the fact that, as the local value of $\RE$ decreases, viscous effects make shear layers diffuse faster so that Kelvin--Helmholtz-like instability modes have not enough time to develop and propagate the cascade toward finer scales (Villermaux, 1998).
 In free shear flows, the {\it mixing transition\/} threshold is consistently found at $\RE_{\lambda}\sim 100$--$140$ (Taylor scale).
It is tempting to conjecture that a similar transition, i.e. a transition from developed turbulence in the featureless regime to chaotically evolving coherent structures in the transitional range, rules the emergence of the laminar-turbulent coexistence as $\RE$ is decreased below $\RE_{\rm t}$.
As compiled by Barkley \& Tuckerman (2007), transitional ranges for a series of different systems are indeed quite similar when using a Reynolds number based on the mean shearing rate.

Another question, likely a difficult one, is about the `optimal' nature of the laminar-turbulent  patterns that are observed in spanwise-extended flows like PCF or PPF.
Extending the minimization of thermodynamic free energies to the weakly out-of-equilibrium domain,  in usual laminar-to-laminar settings, e.g. in Rayleigh--B\'enard convection, pattern selection often happens to achieve the minimization of Ginzburg--Landau potentials that, at least in principle, can be obtained from the primitive equations (Manneville, 1990, chapt.~8--9).
Here, we have to face the problem of predicting the average turbulent fraction, or the typical wavelength and orientation of bands, in circumstances where the overall organization of laminar-turbulent domains is seemingly not directly related to the SSP mechanism at the scale of the MFU.
A Turing-like mechanism with tunable turbulent viscosities would do the job (Manneville, 2012b)  but has not yet been derived from first principles.
Patterning might well be related to the mixing transition as suggested above, but, from a more global perspective, we have no clue for any appropriate potential to optimize, giving the right answer all along the transitional range, from the turbulent-chaotic bifurcation at its uppermost end ($\RE_{\rm t}$) to the chaotic-laminar transition at its lowermost end ($\RE_{\rm g}$).
A decomposition of the flow into mean-flow and coherent structures, on the one hand, and incoherent fluctuations on the other hand, suggests to treat the latter like thermal fluctuations in equilibrium or near-equilibrium thermodynamics.
In the latter case, the most recent theoretical approaches link the derivation of a Gibbs potential to a large-deviation theorem involving small thermal fluctuations (Touchette, 2009).
In the present far-from-equilibrium transition, without detailed balance but a high level of turbulent fluctuations rooted in deterministic Navier--Stokes dynamics, it is not clear how an effective free energy could be defined along similar lines and produce laminar-turbulent patterns by minimization.
 
To conclude, the subcritical transition to$\mskip1mu/\mskip1mu$from turbulence in wall-bounded flows appears as a problem of general interest with the occurrence of pattern formation over a turbulent background.
It should be considered as a testing ground in far-from-equilibrium nonlinear dynamics and complex systems.
Progress in this field might appear of rather conceptual nature, but the tools that have been developed can also serve to a quantitative improvement of transition control in applications, beyond the academic cases that have been examined.
Howsoever biased by the author's interests, the presentation given above is expected to contribute to a better understanding of some main issues in this difficult problem.

\vspace*{2ex}
\noindent {\it Acknowledgements.\/} \quad
Our thanks first go to Prof. Yano for his invitation to present this essay.
Acknowledgements are also due to Y.~Pomeau who introduced us to nonlinear dynamics, to our former collaborators at CEA Saclay, H. Chat\'e, F. Daviaud and other members of the `Instability and Turbulence' group, and to all the participants in the JSPS-CNRS bilateral exchange collaboration {\sc TransTurb},  G. Kawahara \& M. Shimizu in Osaka, T. Tsukahara in Tokyo, Y. Duguet at LIMSI-Orsay, R. Monchaux \& M. Couliou at ENSTA-Palaiseau. Y.~Duguet and a Referee are also thanked for their remarks and suggestions that, we hope, have improved the manuscript.
We also shall not forget our friend, P.~Huerre, who welcomed us  at LadHyX many years ago and provided us with the appropriate resources to do research in our own way.

\end{document}